# Spontaneous Donor Defects and Voltage–Assisted Hole Doping in Beta-Gallium Oxides under Multiple Epitaxy Conditions

Chenxi Nie[1], Kai Liu[1], Chengxuan Ke[1], Xisong Jiang[2], Yifeng He[2], Yonghong Deng[1], Yanhua Yan[2,*], and Guangfu Luo[1,3,*]

[1]*Department of Materials Science and Engineering, Southern University of Science and Technology, Shenzhen 518055, China*

[2]*Shenzhen CAPCHEM Technology Co. Ltd., Shabo Tongfuyu Industry Zone, Pingshan District, Shenzhen 518118, China*

[3]*Guangdong Provincial Key Laboratory of Computational Science and Material Design, Southern University of Science and Technology, Shenzhen 518055, China*

**E-mail: yanyh@capchem.com, luogf@sustech.edu.cn*

## ABSTRACT

Beta-phase gallium oxide (β-$Ga_2O_3$) is prone to the spontaneous formation of donor defects but poses a formidable challenge in achieving high-quality p-type doping, mainly due to its exceptionally low valence band maximum (VBM). In this study, we utilize first-principles computations to investigate the origin of spontaneous donor defects in β-$Ga_2O_3$ grown by three typical techniques: molecular beam epitaxy (MBE), metal organic chemical vapor deposition (MOCVD), and halide vapor phase epitaxy (HVPE). Our findings elucidate that the primary donor defects vary with the growth techniques, specifically $Ga_i^{3+}$ for MBE, $H_i^+$ and $C_{Ga}^+$ for MOCVD, and $(2V_{Ga}+Ga_i+2V_O)^+$ and $Cl_O^+$ for HVPE under unintentionally doped conditions. Employing a theoretically proposed voltage–assisted doping method, we computationally demonstrate that the dominant spontaneous donors can be significantly reduced accompanied by a noticeable increase in acceptors, leading to a stepwise reduction of Fermi level to 0.52, 0.88, and 2.10 eV above VBM for the MOCVD, HVPE, and MBE methods, and a hole concentration of $8.5 \times 10^{17}$, $8.7 \times 10^{11}$, and $2.7 \times 10^{-9}$ $cm^{-3}$, respectively, at room temperature without the use of external dopants. By introducing Mg doping, we further reduce the Fermi level for both the MBE and HVPE experiments.

## 1. INTRODUCTION

Beta-phase gallium oxide ($\beta$-$Ga_2O_3$, *C2/m* space group) is a wide-band-gap semiconductor with several advantages over the traditional semiconductors, such as Si, SiC, and GaN. Specifically, its band gap and breakdown strength are about 4.6–4.9 eV and 8 MV/cm[1], respectively, which enable it to work in ultraviolet, high-power, and radiofrequency fields[2-5]. Additionally, large-size single crystal $\beta$-$Ga_2O_3$ can be readily grown with bulk growth method like the Bridgman method[6] or thin film growth method like the metal-organic chemical vapor deposition (MOCVD)[7]. Its thermal conductivity is relatively low, ranging from 9.5 to 22.5 $Wm^{-1}K^{-1}$ depending on the crystalline direction[8]. As a wide-band-gap semiconductor, $\beta$-$Ga_2O_3$ possesses an extremely low valence band maximum (VBM) of about -9 eV relative to vacuum level[9], which results in easy electron transfer to its conduction band maximum (CBM) but difficult electron transfer from its VBM. Consequently, it is prone to the formation of spontaneous donor defects even without intentional doping, exhibiting electron concentration in the range of $10^{16}$ –$10^{18}$ $cm^{-3}$at room temperature[10-12]. These spontaneous defects potentially reduce the electron diffusivity and largely compensates the p-type doping effect.

However, the defect origins of the spontaneous donor defects are still largely unknown. Earlier studies attributed the unintentional n-type doping to $V_O$[13-15], but was precluded by later theoretical work due to its nature of deep doping. Later, the impurity-related defects, such as of $H_O$, $H_i$[16], Si-related defect[17, 18], and $V_{Ga}$-3H[19], were suggested to be the spontaneous donors. However, these explanations fall short in elucidating the generality of spontaneous n-type doping across different growth methods, such as molecular beam epitaxy (MBE)[20], MOCVD[10], halide vapor phase epitaxy (HVPE)[11], floating zone method[15], Czochralski method[17], edge-defined film-fed growth method[18, 19], and Verneuil method[13]. For instance, the MBE is known for its low impurity levels. Since the experimental identification of defect composition is extremely difficult and most theoretical studies did not explore the defects under the specific growth conditions of $\beta$-$Ga_2O_3$, the defect origins of spontaneous donors remain to be answered.

Besides the existence of abundant spontaneous donors, the high-quality p-type doping has not been realized in $\beta$-$Ga_2O_3$ yet. Previous N doping[21] or Mg doping[22] through magnetron sputtering was reported to achieve a very weak p-type doing. Recent experiments through oxidization of GaN under oxygen ambient led to N-doped $\beta$-$Ga_2O_3$ films[23, 24], realizing a hole concentration of $4.8 \times 10^{15}$–$1.6 \times 10^{16}$ $cm^{-3}$ and a hole mobility of 13.8–23.6 $cm^2V^{-1}s^{-1}$ under room temperature. Interestingly, there were also reports that p-type $\beta$-$Ga_2O_3$ were achieved without intentional doping. For instance, the pulsed laser deposition grown samples exhibited a hole concentration about $2 \times 10^{13}$ $cm^{-3}$ and a hole mobility of ~0.2 $cm^2V^{-1}s^{-1}$ at room temperature[25]; $\beta$-$Ga_2O_3$ films grown by MOCVD followed by an annealing process was reported to exhibit a hole concentrations of ~$3 \times 10^{14}$ $cm^{-3}$ at room temperature[26]. Theoretically, previous density-functional-theory (DFT) studies predicted a high charge transition level of 0/- for $Mg_{Ga2}$, ~1 eV above VBM[27, 28], which was thought too high to induce an efficient p-type doping. This prediction, however, is significantly higher than a recent experimental value of 0.65 eV based on the electron paramagnetic resonance (EPR)[29]. To decode these puzzling reports and potentially address the p-type doping challenge for β-$Ga_2O_3$, in-depth studies with high accuracy are highly desirable.

In this article, we systematically examine all the major point defects under the MBE, MOCVD, and HVPE growth conditions based on first-principles calculations and reveal the diverse spontaneous donors in β-$Ga_2O_3$. Furthermore, we apply a recently proposed voltage–assisted doping approach[30] to shift the band edges and computationally demonstrate significant reduction of all the spontaneous donors and dramatic increase of desirable acceptors under all the typical growth conditions.

## 2. METHODS

### *2.1 Computational details of density functional theory calculations*

All the first-principles calculations are based on the density functional theory (DFT), as implemented in the Vienna Ab-initio Simulation Package[31]. The plane-wave energy cutoff is set to 485 eV for all calculations except the molecular dynamics (MD), which utilize a value of 404 eV. We adopted the following projector augmented wave (PAW)[32] pseudopotentials: Ga_d_GW($3d^{10}4s^24p^1$) for Ga, O_s_GW($2s^22p^4$) for O, C_GW($2s^22p^2$) for C, H_GW($1s^1$) for H, Cl_GW($3s^23p^5$) for Cl, Mg_GW($3s^2$) for Mg. Because a reliable defect thermodynamic calculation requires an accurate description of the geometrical structure, band gap, and energy, we utilize the Heyd-Scuseria-Ernzerhof (HSE06) hybrid functional[33] and the Hubbard U approach for the $3d$ electrons in gallium, with an effective $U_{eff}$ of 2.7 eV. As elaborated in Section I of the Supplementary Information, this approach gives an overall optimal description of the structure, band gap, and formation enthalpy of β-$Ga_2O_3$, outperforming the typical method of setting the hybrid mixing parameter $\alpha$ to 0.30-0.35.[16, 27, 28, 34-40] Note that previous studies showed that the HSE06+U method predicted the band gaps and structures of II-VI semiconductors noticeably better[41]. The $k$-point sampling spacing in Brillouin zone is 2π/80 $Å^{-1}$ for the density of states calculations, 2π/15 $Å^{-1}$ for the MD calculations, and about 2π/25 $Å^{-1}$ for the rest. For defect energy calculation, a 1 × 3 × 2 supercell with 120 atoms is adopted. To reduce the image-charge interactions among supercells, the Freysoldt-Neugebauer-Van (FNV)[42] corrections are employed, with a dielectric constant of 11.16[43].

### *2.2 Voltage-assisted doping approach*

We utilize a recently developed theoretical method, voltage-assisted doping approach, to tune the defect thermodynamics[30]. In this approach, a proper external voltage is applied to the semiconductors during growth or doping processes, so one can induce band bending and charge accumulation that reverses the defect thermodynamics. For $\beta$-$Ga_2O_3$, a positive external voltage can bring down band edges and induce electron accumulation, which consequently render the donors thermodynamically unstable and the acceptors stable. For a defect $A_B^q$ with element $A$ occupying site $B$ and a charge state of $q$, its defect formation energy under a band bending value of $U_G$ can be expressed as Eq. (1).

$$E_f(A_B^q, E_{FG}, T_G) \equiv E_{tot}(A_B^q) + E_{FNV}(A_B^q) + \mu_B(T_G) - \mu_A(T_G) + q[E_{FG}(T_G) + E_{VBM}(\text{perfect}, T_G) - U_G] - E_{tot}(\text{perfect}), \quad (1)$$

where $E_{tot}$ represents the total energy, $E_{FNV}$ is a finite-size-correction[42], $\mu_B(T_G)$ and $\mu_A(T_G)$ are chemical potentials under experiment conditions, including partial pressure and growth temperature $T_G$ (see details in the notes of Table 1), $E_{FG}(T_G)$ is the Fermi level at growth temperature, and $E_{VBM}$(perfect, $T_G$) is the VBM energy of perfect structure under growth temperature. In the absence of an applied external voltage, $U_G$ equals zero. To obtain the VBM energy under high growth temperatures, we carry out MD simulations and calculate the VBM energy for sampled structures (Section II of the Supplementary Information).

With the defect formation energy expression in Eq. (1) and the charge neutrality condition in Eq. (2), the specific Fermi level under growth conditions ($E_{FG}$) and defect concentration ($c_{defect}$) can be self-consistently determined.

$$n_h(T_G, E_{FG}) - n_e(T_G, E_{FG}) + \sum c_{defect}(T_G, E_{fG}, E_{FG}) * q(E_{FG}, T_G) = 0, \quad (2)$$

$$c_{defect} = c_{site} \exp(-E_f / k_B T_G), \quad (3)$$

where the free hole concentration $n_h$ and electron concentration $n_e$ at temperature $T$ are calculated based on the temperature-dependent density of states, $DOS(T, E)$, and the Fermi-Dirac distribution; $c_{site}$ is the maximum site concentration for a defect in crystal.

After the growth, the crystal cools down to the room temperature $T_R$ and the external voltage is removed. The defect concentrations are supposed to be largely frozen (see Section III of the Supplementary Information), but the defect charges and free charge carriers can relax to a new equilibrium. As a result, the Fermi level at room temperature, $E_{FR}$, is determined with the new charge neutrality condition in Eq. (4).

$$n_h(T_R, E_{FR}) - n_e(T_R, E_{FR}) + \sum c_{defect}(T_G, E_{fG}, E_{FG}) * q(E_{FR}, T_R) = 0 \quad (4)$$

Additional information on the voltage-assisted doping approach can be found elsewhere[30].

While practical methods for integrating this approach with existing epitaxy techniques have yet to be developed, two studies provide valuable insights. In one study, two electrodes were positioned above and below the substrate during the growth of $\beta$-$Ga_2O_3$, resulting in notable changes in growth rate, surface morphology, and other material properties[44]. In another study, external voltage was used as a post-processing technique. The authors applied an external voltage to a $SrCoO_x$ thin film under an oxygen atmosphere and successfully tuned the concentration of oxygen vacancy[45]. This method avoids the modification of growth equipment, making it easily applicable for controlling certain defects in other thin films. To mitigate the risk of electrochemical reactions caused by the external voltage, we recommend placing an insulating film above the electrodes to block charge transfer [30]. Furthermore, one can apply this approach to epitaxial thin films, where a low voltage can induce substantial band bending without triggering undesirable electrochemical reactions.

## 3. RESULTS AND DISCUSSION

### *3.1 Experimental growth conditions of β-$Ga_2O_3$ and potential point defects*

We closely examine the defects under three typical growth conditions of β-$Ga_2O_3$ without intentionally doping: the MBE, MOCVD, and HVPE, as schematically shown in Figure 1. For the MBE, the growth temperature is around 823 K and the oxygen partial pressure is ~$1.3 \times 10^{-3}$ Pa[46, 47]. The impurities in MBE samples can be as low as $10^{13}$ $cm^{-3}$ and are typically disregarded[48]. The MOCVD utilizes a growth temperature of 1023 K and the gallium source is trimethylgallium (TEGa) or trimethylgallium (TMGa); the oxygen partial pressure is ~$6.1 \times 10^3$ Pa[49, 50]. The C and H are typical impurities in this method. The HVPE operates at a growth temperature of about 1273 K, and the gallium source is GaCl, which is the reaction product of $Cl_2$ and Ga metal. Therefore, chlorine is a typical impurity in the final products[51, 52]. The oxygen partial pressure is at the level of $5.1 \times 10^2$ Pa. The detailed growth conditions and chemical potentials are summarized in Table 1.

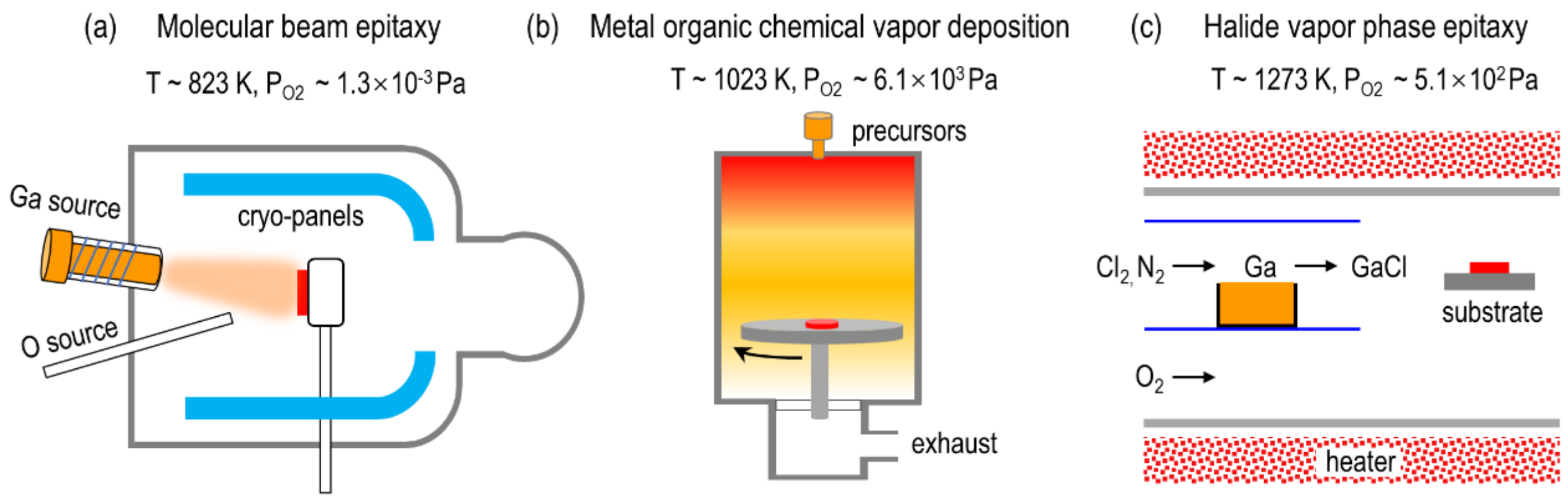

Figure 1. Schematic device of the MBE(a), MOCVD(b), and HVPE(c), together with the critical growth conditions.

Table 1. Growth temperature, oxygen partial pressure, gallium source and its partial pressure, impurity concentration, and its chemical potential for the typical MBE[46, 47], MOCVD[49, 50], and HVPE[51, 52] of β-$Ga_2O_3$. The chemical potentials of the Ga-rich and O-rich conditions are also provided for comparison.

| | Ga-rich* | MBE | MOCVD | HVPE | O-rich* |
|---|---|---|---|---|---|
| $T_G$ (K) | | 823 | 1023 | 1273 | |
| $P_{O_2}$ (Pa) | | ~1.33 × $10^{-3}$ | ~6.08 × $10^{3}$ | ~5.07 × $10^{2}$ | |
| Ga source, $P_{Ga}$ (Pa) | | Ga vapor, ~2.0 × $10^{-4}$ | TEGa gas, 2.7-10 | GaCl gas, 1.0 × $10^{2}$ | |
| $\mu_O$ (eV)† | -10.69 | -8.37 | -8.10 | -8.59 | -6.90 |
| $\mu_{Ga}$ (eV)‡ | -3.27 | -3.41 | -7.69 | -7.27 | -8.95 |
| $C_{impurity}$ ($cm^{-3}$) | | – | C: ~2.0 × $10^{16}$<br>H: ~1.2 × $10^{17}$ | Cl: ~1.5 × $10^{16}$ | |
| §$\mu_{impurity}$ (eV) | | – | $\mu_C$: -12.76<br>$\mu_H$: -6.22 | $\mu_{Cl}$: -4.30 | |

†The translational, vibrational, and rotational entropies are included under experimental conditions.

‡The chemical potential of $\mu_{Ga}$ is obtained using the equilibrium condition of $2\mu_{Ga} + 3\mu_O = \mu_{Ga_2O_3}$ under experimental conditions, except for the MBE, in which the $\mu_{Ga}$ is calculated directly from the pressure and temperature of gallium vapor.

§The chemical potentials of impurities are determined in a way that the impurity concentrations are consistent with the experimental values.

*The chemical potentials of $\mu_{Ga}$ in the Ga-rich and $\mu_O$ in the O-rich equal to the total energy per atom of gallium metal and oxygen molecule, respectively. The $\mu_{Ga}$ in O-rich and $\mu_O$ in Ga-rich are obtained through the equilibrium condition of $2\mu_{Ga} + 3\mu_O = \mu_{Ga_2O_3}$, where $\mu_{Ga_2O_3}$ is energy per formula of gallium oxide bulk. No temperature or partial pressure effects are considered here.

In total, we have examined 34 defects, including 23 intrinsic defects and 11 impurity ones, as listed in Table 2. Seven complex defects involving three or more point defects are examined based on previous theoretical and/or experimental observations[36, 40, 53-55]. The detailed process of selecting defect types is provided in Section IV of the Supplementary Information. All the optimized defect structures are also provided in the Supplementary Information.

Table 2. Thirty-four defects examined for three types of growth methods. Note that β-$Ga_2O_3$ possesses two inequivalent Ga positions, $Ga_1$ and $Ga_2$, three inequivalent oxygen positions, $O_1$, $O_2$ and $O_3$, and three interstitial positions, named as 1, 2 and 3, as shown in Figure 2(i).

| | MBE | MOCVD | HVPE |
|---|---|---|---|
| Intrinsic defects | $V_{O1}$, $V_{O2}$, $V_{O3}$, $V_{Ga1}$, $V_{Ga2}$, $Ga_{O1}$, $Ga_{O2}$, $Ga_{O3}$, $O_{Ga1}$, $O_{Ga2}$, $O_{i1}$, $O_{i3}$, $Ga_{i1}$, $Ga_{i3}$, $Ga_{i3}$+$O_{i1}$, $V_{O2}$+$O_{i1}$, $V_{O2}$+$V_{Ga2}$, $V_{Ga1}$+$V_{Ga2}$+$Ga_{i1}$[53, 54], 2$V_{Ga1}$+$Ga_{i2}$[53, 54], 2$V_{Ga1}$+$Ga_{i3}$[53, 54], 2$V_{Ga1}$+$Ga_{i2}$+$V_{O1}$[36, 40], 2$V_{Ga1}$+$Ga_{i2}$+2$V_{O1}$[36], 2$V_{Ga1}$+$Ga_{i3}$+$V_{O3}$[36] | | |
| Impurity defects | – | $C_{Ga1}$, $H_{Ga1}$, $H_{i1}$, $H_{i2}$, $H_{Ga1}$+$C_{Ga1}$, $H_{Ga1}$+$H_{i2}$ $C_{Ga1}$+$V_{Ga2}$, 2$H_{Ga1}$+$Ga_{i2}$[55] | $Cl_{O1}$, $Cl_{Ga1}$, $Cl_{O1}$+$V_{Ga2}$ |

*3.2 Explanation of spontaneous n-type doping in β-$Ga_2O_3$*

The defect formation energy under the growth temperatures is shown in Figure 2 and the results are divided into intrinsic point defects, intrinsic pair defects, and impurity defects for easy visualization. For the MBE, the self-consistent Fermi level locates at 2.30 eV under the growth temperature of 823 K and the dominant defects are $Ga_i^{3+}$ and $V_O^{2+}$, as illustrated in Figure 2(a) and (b). These findings remain the same for MBE growth employing $O_3$ as the oxygen source, as detailed in Section V of the Supplementary Information. The impurity defects are neglected for MBE, because of its high purity growth conditions. For the MOCVD [Figure 2(c)-(e)], the self-consistent Fermi level is at 2.13 eV at the growth temperature of 1023 K. Because of the significantly reduced chemical potential of gallium relative to the MBE (Table 1), the most abundant intrinsic defects change to $(2V_{Ga1}+Ga_{i2}+V_{O1})^-$, $(2V_{Ga1}+Ga_{i2}+2V_{O1})^+$, and $(2V_{Ga1}+Ga_{i3}+V_{O3})^-$. Because of the organic precursors, the hydrogen- and carbon-related defects, $H_{i2}^+$, $H_{i1}^+$and $C_{Ga1}^+$, become dominant in MOCVD. For the HVPE, the self-consistent Fermi level is 2.02 eV at the growth temperature of 1273 K [Figure 2(f)-(h)). The dominant intrinsic defects are $(2V_{Ga1}+Ga_{i2}+2V_{O1})^+$, $(2V_{Ga1}+Ga_{i2}+V_{O1})^-$, $(2V_{Ga1}+Ga_{i3}+V_{O3})^-$, and $V_O^{2+}$. Given the GaCl precursor, the Cl-related defect, $Cl_{O1}^+$, are unavoidable and exist in a significant amount.

After cooling down to room temperature, the defect concentrations are supposed to be largely frozen (see Section III of the Supplementary Information), while the defect charges and free charge carriers can relax to a new equilibrium. Figure 3 summarizes the charge concentrations of all the major defects at room temperature under unintentionally doped conditions. Among the different growth methods, β-$Ga_2O_3$ grown by the MBE has the lowest donor charge concentrations of $1.5\times10^{15}$ $cm^{-3}$. MOCVD and HVPE samples exhibit higher donor charge densities of $5.5\times10^{16}$ and $1.7\times10^{17}$ $cm^{-3}$, respectively. These results clearly indicate that the defect origins of spontaneous n-type doping in β-$Ga_2O_3$ strongly depend on the growth method: intrinsic $Ga_{i3}^{3+}$ for MBE, impurities $H_{i2}^{+}$, $H_{i1}^{+}$, and $C_{Ga1}^{+}$ for MOCVD, and intrinsic $(2V_{Ga1}+Ga_{i2}+2V_{O1})^{+}$ and impurity $Cl_{O1}^{+}$ for HVPE.

Note that $H_i$ is expected to be quite mobile, which would lead to a reduced concentration in experiments; In this context, $C_{Ga1}^{+}$ would play a more significant role in the spontaneous doping process. The structure of $(2V_{Ga1}+Ga_{i2}+2V_{O1})^{+}$ can be interpreted as a combination of two oxygen vacancies with an experimentally identified defect $2V_{Ga1}+Ga_{i2}$[36, 53]. Despite having fewer defects than the MOCVD and HVPE samples, the electron mobilities of the MBE samples are found to be the lowest among the three growth methods at the same electron concentration[56]. This seemingly paradoxical phenomenon is likely due to the higher charge state of $Ga_{i3}^{3+}$, which exhibits stronger Coulombic interactions with charge carriers, leading to reduced mobility.

Under the room temperature, the self-consistent Fermi levels are equilibrated to 3.89 eV for MBE and 4.01 eV for MOCVD and HVPE samples, respectively (Figure 3). Given that the band gap of β-$Ga_2O_3$ is ~4.46 eV at room temperature, all the samples exhibit typical n-type trait without intentional doping, and the free electron concentrations are $1.5\times10^{15}$, $1.4\times10^{17}$, and $1.5\times10^{17}$ $cm^{-3}$ for the MBE, MOCVD, and HVPE, respectively, in good agreement with the experimental values under similar growth conditions (Figure 3)[10, 11, 20]. We have also made efforts to correlate the charge transition levels of the dominant defects in Figure 3 with experimental results, achieving reasonable consistency (Section VI of the Supplementary Information). However, these comparisons remain at a preliminary stage due to the discrepancies among experimental data, which could arise from the existence of impurities and varying experimental temperature[Phys. Rev. B 105, 115201 (2022)].

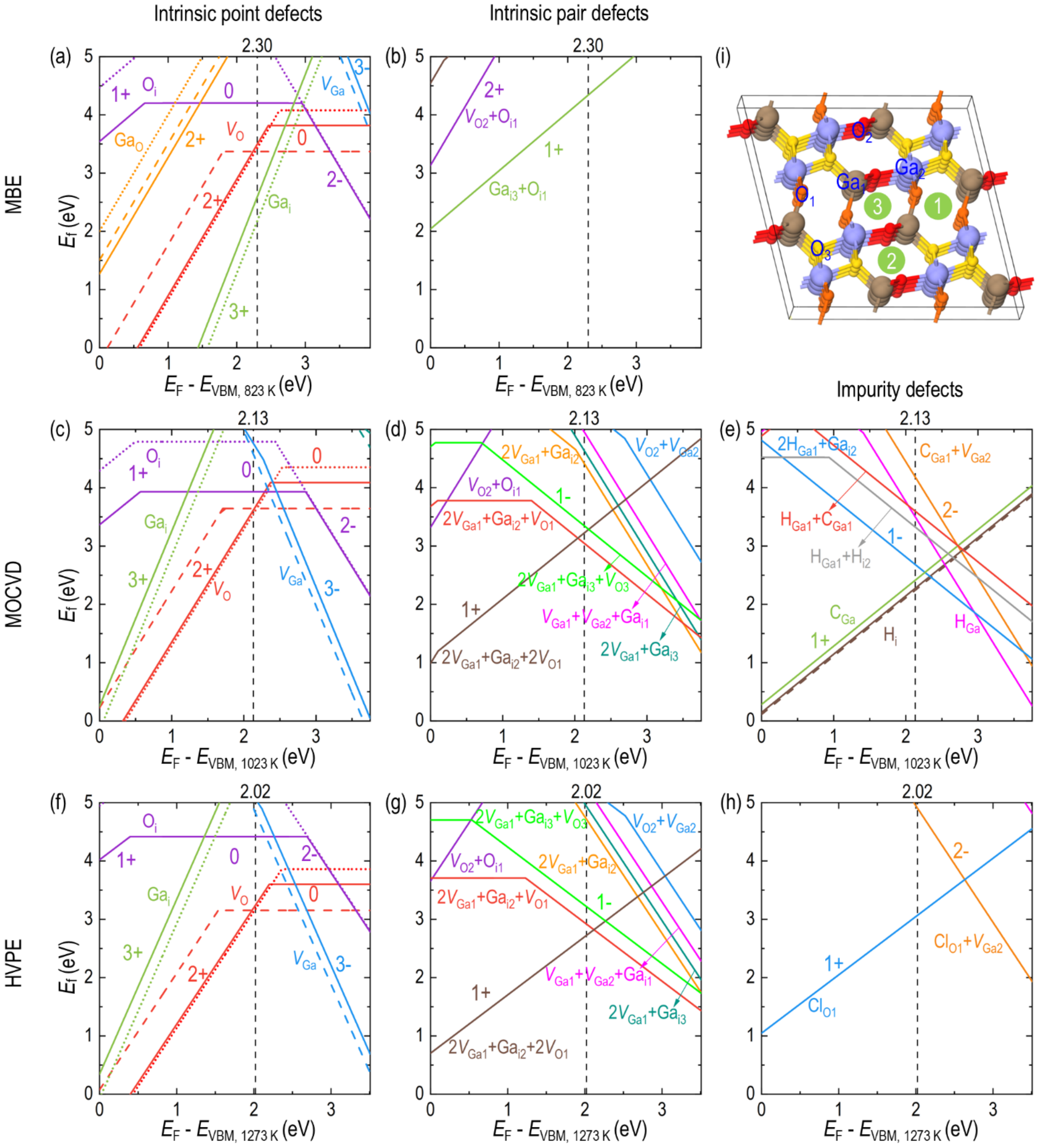

Figure 2. Defects formation energy for the (a-b) MBE, (c-e) MOCVD, and (f-h) HVPE under their respective growth temperatures; (i) geometrical structure of β-$Ga_2O_3$ and the inequivalent atomic sites. The vertical dash line indicates the self-consistent Fermi level $E_{FG}$. Defect charge states are labeled besides the slopes in panel (a). The solid, dashed, and short curves represent defects occupying the first, second, and third inequivalent positions in β-$Ga_2O_3$, respectively. Parallel lines within and across different panels represent the same charge states.

We have also examined the impacts of growth temperature on the defect formation (Section VII of the Supplementary Information). Our findings indicate that the dominant defect types remain the same under different temperatures. However, the defect and carrier concentrations slightly increase or decrease depending on the growth techniques. This complexity arises because growth temperature influences the defect concentration both explicitly (Eq. 2) and implicitly through the defect formation energy.

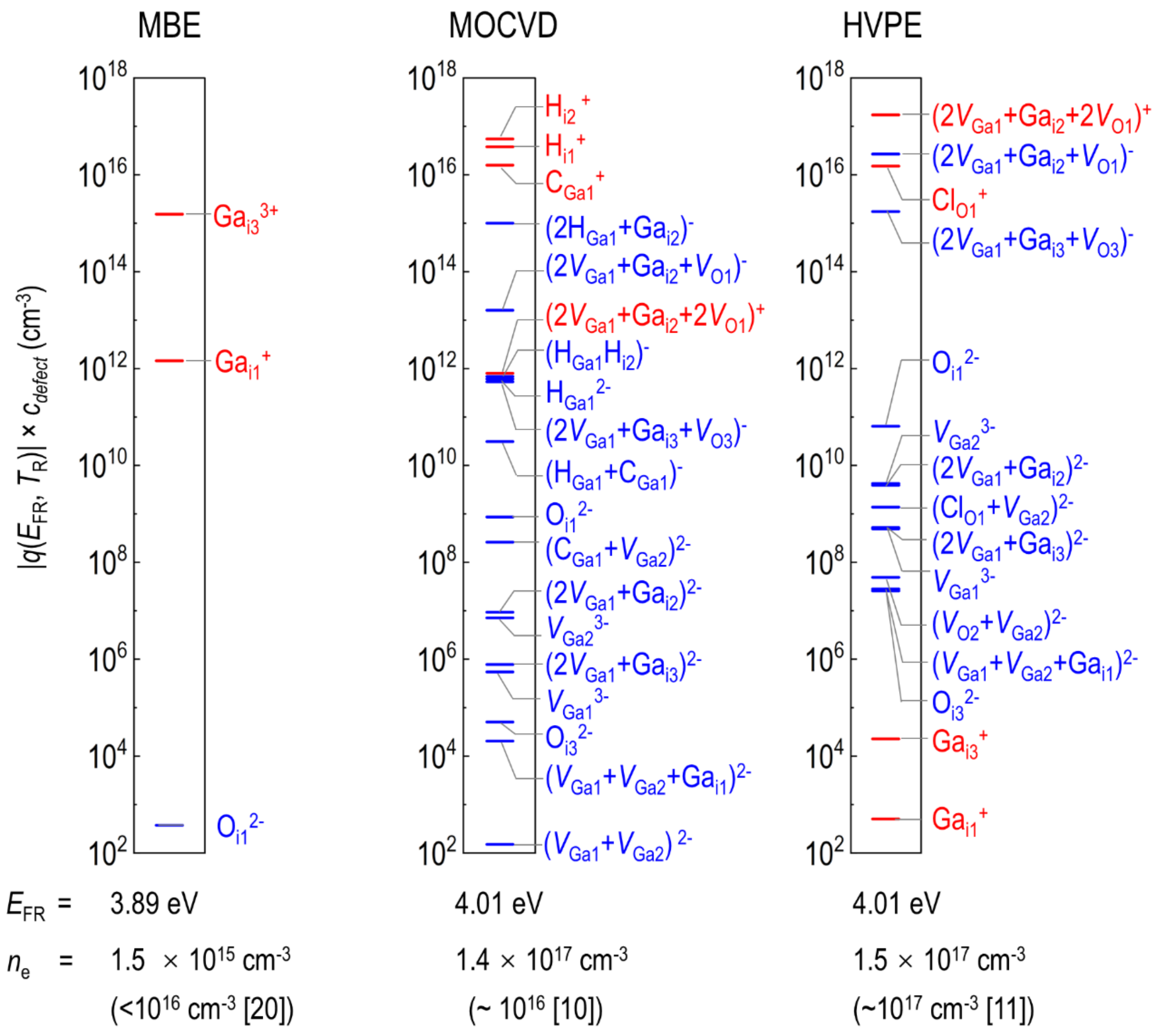

Figure 3. Charge concentrations of defects, self-consistent Fermi level ($E_{FR}$), and free electron concentration ($n_e$) at room temperature for the three growth methods. The experimental $n_e$ is shown in paratheses.

*3.3 Application of voltage-assisted doping approach to β-$Ga_2O_3$*

To reverse the defect thermodynamics in β-$Ga_2O_3$, namely making the stable donors unstable and unstable acceptors stable, we computationally explore a voltage-assisted doping approach in the three growth methods. This approach leverages an external voltage to induce a band bending in the target material, as reversely impacts the formation energies of donor and acceptor defects. Further details can be found in the Methods section or elsewhere[30]. As shown in Figure 4a, with the decreasing band bending $U_G$, $E_{FR}$ in the MBE sample first gradually decreases from 3.9 to 3.2 eV. At $U_G \sim -0.8$ eV, $E_{FR}$ suddenly drops to 2.8 eV and remains almost constant till $U_G \sim -1.7$ eV, where it drops slightly followed by an abrupt decrease to 2.1 eV at $U_G \sim -1.8$ eV. The above phenomenon is caused by the defect charge concentration change in Figure 4d. In the initial stage, the concentration of the major donor $Ga_{i3}^{3+}$ is orders' more than the acceptors and it decreases almost exponentially with increasing $U_G$, as induces the quasi-linear decrease of $E_{FR}$. Meanwhile, the concentration of acceptor defect $O_{i3}^{2-}$ starts an exponential increase beginning at $U_G \sim -0.4$ eV and it surpasses that of $Ga_{i3}^{3+}$ at $U_G \sim -0.8$ eV. After this critical point, $E_{FR}$ is pinned to the 0/-2 donor level of $O_{i3}^{2-}$ at 2.8 eV, corresponding to the first terrace in Figure 4a. In the range of $U_G \sim -0.8$ to $\sim -1.7$ eV, the charge concentrations of major donor defect ($V_{O3}^{2+}$) and major acceptor defect ($O_{i3}^{2-}$) remain almost equal, which explains the first $E_{FR}$ terrace. When $U_G$ exceeds $-1.7$ eV, the acceptor defect $V_{Ga2}^{3-}$ successively surpasses $V_{O3}^{2+}$ and $V_{O1}^{2+}$, causing $E_{FR}$ eventually pinned to the donor level of $V_{O2}^{2+}$ at 2.1 eV, which corresponds to the final terrace in Figure 4a. Overall, the voltage-assisted doping approach dramatically decreases the spontaneous donor defects by orders in MBE and the $E_{FR}$ can be tuned to 2.1 eV (slightly below the midgap) with a low hole concentration ($2.7\times10^{-9}$ cm$^{-3}$). No further reduction of $E_{FR}$ is realized here, majorly because the formation energy of acceptors in MBE method is too high to be significantly tuned in the reasonable range of band bending, which is less than half of the band gap[30].

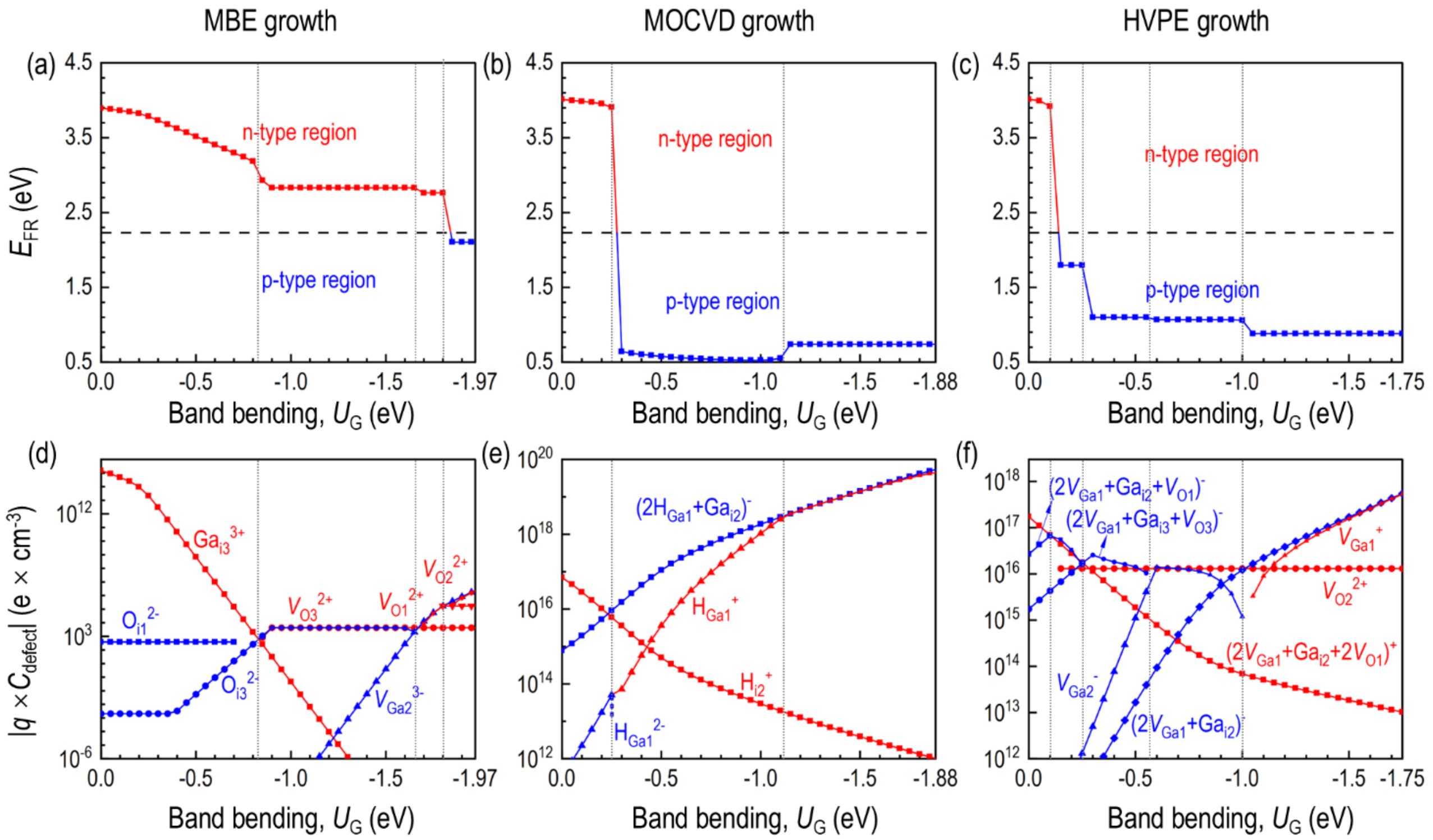

Figure 4. Room temperature Fermi level, $E_{FR}$, versus the band bending $U_G$ for the (a) MBE, (b) MOCVD, and (c) HVPE, respectively. The charge concentrations of dominant defects are shown in (d-f), where the defects with zero charge are invisible.

The $E_{FR}$ terraces in Figure 4a reveal one interesting feature of defects: when the Fermi level is pinned to a defect level, the defect can exhibit one charged state and one neutral state simultaneously. For example, in the first terrace of MBE, the charge concentration of $O_{i3}^{2-}$ remains constant, when the total concentration of $O_{i3}$ continues to increase with $U_G$; the extra defects are found to exist in the form of $O_{i3}^0$. Such phenomenon is observed in all $E_{FR}$ terraces of Figure 4(a-c). This finding reminds us that the actual defect concentration in experiments could be significantly more than that calculated from the free charge carriers for a system with pinned Fermi level.

For MOCVD, the $E_{FR}$ also shows a dramatical decline from 4.0 to 0.6 eV at $U_G \sim -0.3$ eV (Figure 4b), as is caused by the surpass of the dominant acceptor $(2H_{Ga1}+Ga_{i2})^-$ over the dominant donor $H_{i2}^+$ (Figure 4e). Such abrupt change transforms the material to p-type doping region. With decreasing $U_G$, the concentration of $(2H_{Ga1}+Ga_{i2})^-$ gradually rises, progressively leading $E_{FR}$ towards the 0/- energy level of $(2H_{Ga1}+Ga_{i2})^-$ at 0.40 eV. Interestingly, a slight jump at $U_G \sim -1.2$ eV increases $E_{FR}$ to 0.74 eV, which is caused by the surpassing of $H_{Ga1}^+$ concentration over that of $(2H_{Ga1}+Ga_{i2})^-$. This phenomenon seems controversial, given that a negative $U_G$ should increase the formation energy of a donor and decreases its concentration according to Eqns. 1. The physics behind is that $H_{Ga1}$ is a bipolar defect, which behaves as an acceptor with -2 charge state at the growth temperature but a donor with +1 charge state at room temperature when $E_{FR}$ is less than 0.74 eV. For MOCVD, the lowest $E_{FR}$ is 0.52 eV realized at $U_G \sim -1.0$ eV, which corresponds to a hole concentration of $8.5 \times 10^{17}$ cm$^{-3}$. Note that no external dopants are introduced here and the p-type doping is achieved through the intrinsic defects and/or the impurities unavoidable in the experiments.

The $E_{FR}$ in HVPE exhibits a unidirectional decline with the decreasing $U_G$ (Figure 4c). Its abrupt decreases appear at $U_G$ of $\sim -0.1$, $-0.3$, $-0.6$ and $-1.0$ eV, where the charge concentration of a dominant acceptor surpasses that of a dominant donor (Figure 4f). At $U_G \sim -1.0$ eV, the spontaneous donors are dramatically reduced to $\sim 10^{16}$ cm$^{-3}$. Above $U_G \sim -1.0$ eV, the $E_{FR}$ is pinned to 0.88 eV, which corresponds to a hole concentration of $8.7\times10^{11}$ cm$^{-3}$.

*3.4 Mechanism of Fermi-level tunning by voltage-assisted doping approach*

To further illustrate the mechanisms and relationships in Figure 4, we propose two simplified examples in Figure 5. The first example consists of one donor defect $D^+$ and one acceptor defect A (Figure 5a), which exhibits a charge transition level of 0/- at $E_A^{0/+}$. Initially, the concentration of $D^+$ is more than that of $A^-$, a situation corresponding to the n-type doping. With a more negative $U_G$, $E_{FR}$ gradually approaches the intersection of the two defects (VBM and CBM also shift to the right by $U_G$). Meanwhile, the concentration of $D^+$ decreases and that of $A^-$ increases (Figure 5b), which leads to a quasi-linear decrease of $E_{FR}$. At the critical point of $U_G = U_1$, the charge concentrations of decreased $D^+$ and increased $A^-$ become equal, resulting in a sudden drop of $E_{FR}$ to the acceptor level of $A^{0/-}$ at room temperature (Figure 5c), namely, $E_A^{0/-}$ + ($E_{VBM,TG} - E_{VBM,TR}$). Beyond the critical point, the concentration of $A^0$ rapidly increases from zero, while the concentrations of $A^-$ and $D^+$ balance with each other and both gradually decrease. In the range of $U_G <$ $U_1$, defect A exhibits the coexistence of charge states $A^0$ and $A^-$, namely that defect A is partially ionized.

The second example involves a third bipolar defect B, which exhibits two energy levels of +/0 and 0/2- within the band gap (Figure 5d). At $U_G > U_1$, the concentration of $D^+$ is greater than both $A^-$ and $B^{2-}$, and the $E_{FR}$ declines gradually, similar to the first example. For $U_2 < U_G < U_1$, the defect concentration of A dominates and the neutral $A^0$ emerges. Meanwhile, $E_{FR}$ drops to the energy level of $A^{0/-}$. At the critical point of $U_G = U_2$, the concentration of $B^+$ surpasses that of $A^-$ (Figure 5e) and induces a sudden $E_{FR}$ withdrawal to the energy level of $B^{+/0}$ (Figure 5f). In the range of $U_G < U_2$, the neutral $B^0$ emerges, and its concentration increases rapidly, while $A^-$ and $B^+$ maintains a balance.

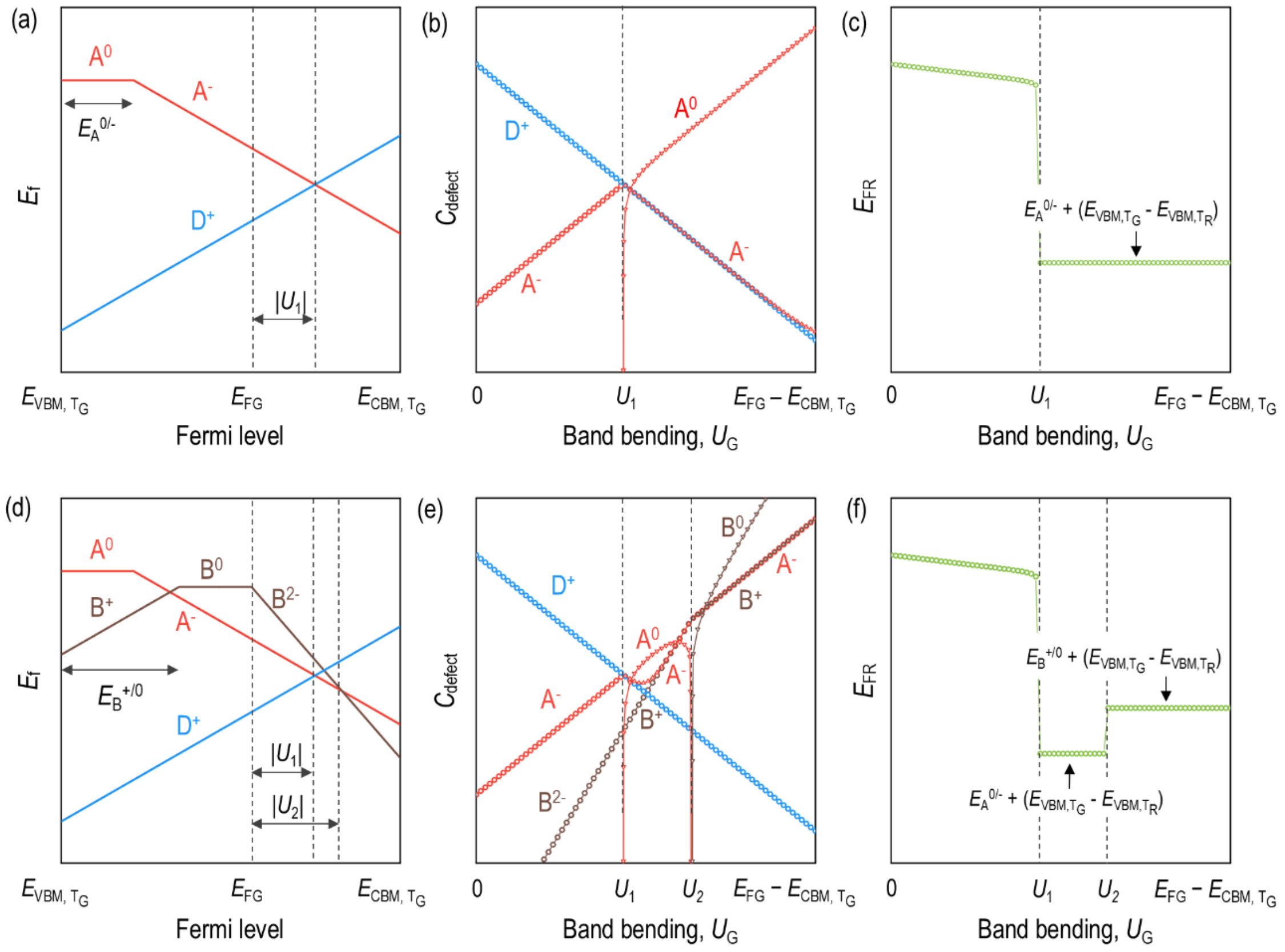

Figure 5. Schematic defect formation energy $E_f$ at growth temperature, defect concentration $C_{defect}$, and the room-temperature Fermi level $E_{FR}$ for cases (a-c) without and (d-f) with bipolar defect. Note that $U_G$ is negative in this case.

### *3.5 Discussions on the p-type doping of β-$Ga_2O_3$*

After reducing/increasing the donor/acceptor concentration, the lowest acceptor energy level becomes the fundamental factor limiting the lowest $E_{FR}$. Therefore, we list all the energy levels at room temperature below 2 eV in Figure 6, including that of the dopant defect $Mg_{Ga2}$. Among all the defects, $Mg_{Ga2}$ and $2H_{Ga1}$+$Ga_{i2}$ exhibit the lowest acceptor levels at ~0.39 and 0.40 eV, respectively, while $2V_{Ga1}$+$Ga_{i2}$ exhibits the lowest acceptor level at 0.79 eV among intrinsic defects. Because of its heavy effective hole mass of

~7.6 $m_e$ and wide band gap, β-$Ga_2O_3$ can support efficient ionization of dopants with acceptor levels up to 0.60 eV at room temperature (Section VIII of the Supplementary Information). To increase the hole concentrations of β-$Ga_2O_3$ for MBE and HVPE, we explore the Mg doping together with the voltage-assisted approach. As shown in Figure 7, a chemical potential of Mg greater than −4.35 eV for MBE or −2.64 eV for HVPE can reduce the $E_{FR}$ to 0.39 eV and realize a hole concentration of $1.5\times10^{20}$ $cm^{-3}$ at room temperature, suggesting its potential as an effective acceptor for β-$Ga_2O_3$. Certainly, the current method could also benefit from other strategies for modulating electronic properties, such as the bandgap narrowing effects and using alloying to shift the band edges and tune defects.[57]

It is noteworthy that a consensus has yet to be reached regarding the energy level of $Mg_{Ga2}^{0/-}$, as outlined in Table S4 (Section IX of the Supplementary Information), which summarizes the major experimental and theoretical findings. Our HSE06+U calculations predict an energy level of $E_{VBM(300\ K)}$ + 0.39 eV, which aligns with an experimental value of about $E_{VBM(240\text{-}260\ K)}$ + 0.65 eV based on electron paramagnetic resonance[29, 58]. However, the prediction based on HSE ($\alpha$ = 0.32) is about $E_{VBM(0\ K)}$ + 1.0 eV according to both literature[27, 28] and our own calculation (or $E_{VBM(300\ K)}$ + 0.84 eV with consideration of temperature effects), which also receives experimental support [59, 60]. One reason for this discrepancy is that the experimental identification of defects heavily relies on comparison with theoretical predictions. While it may be premature to draw conclusions on this matter, it is noteworthy that certain native defects, such as $2V_{Ga1}$+$Ga_{i3}$+$V_{O3}$, indeed exhibit both a high concentration (Figure 3) and acceptor energy levels around the experimental value of ~1.0 eV (Figure 6).

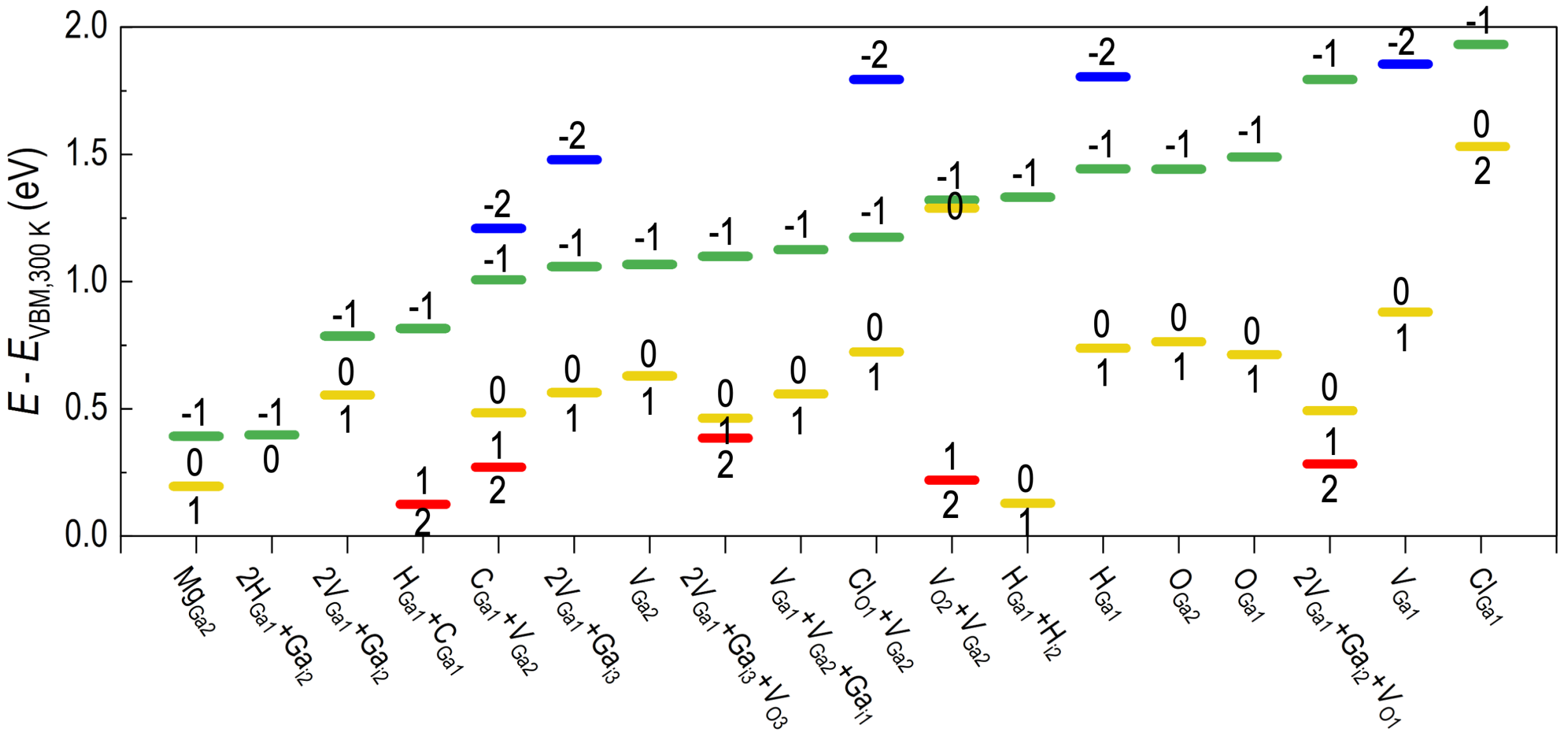

Figure 6. Charge transitional levels of the examined defects below $E_{VBM}$ + 2 eV at room temperature. Number below or above an energy level represents the defect charge in the region immediately below or above the energy level.

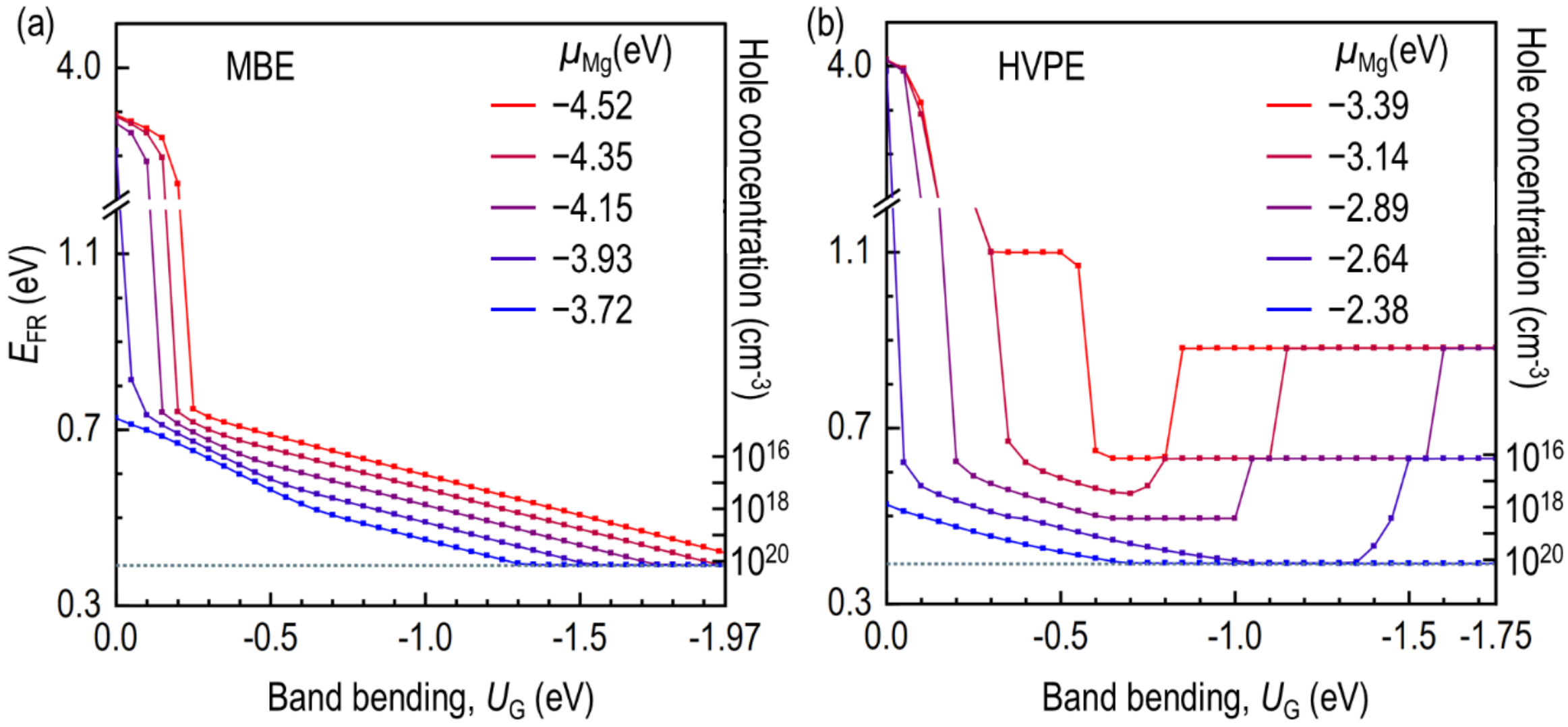

Figure 7. Room temperature Fermi level, $E_{FR}$, versus the band bending $U_G$ for Mg-doped β-$Ga_2O_3$ under the growth conditions of (a) MBE and (b) HVPE. The five increasing chemical potentials of Mg in the legends, respectively, correspond to Mg concentrations of $10^{14}$, $10^{15}$, $10^{16}$, $10^{17}$, and $10^{18}$ $cm^{-3}$ at $U_G = 0$ eV.

## 4. CONCLUSION

In summary, we computationally studied the defects in β-$Ga_2O_3$ under the MBE, MOCVD, and HVPE conditions based on the HSE06+U method and apply the voltage-assisted-doping approach to tune the defect thermodynamics. Through a careful examination of the defect formations under growth temperatures and the charge rebalance at room temperature, we explain the spontaneous n-type doping in β-$Ga_2O_3$, which is majorly induced by $Ga_{i3}^{3+}$ for MBE, $H_{i2}^{+}$, $H_{i1}^{+}$, and $C_{Ga1}^{+}$ for MOCVD, and $(2V_{Ga1}+Ga_{i2}+2V_{O1})^{+}$ and $Cl_{O1}^{+}$ for HVPE under unintentionally doped conditions. With the voltage-assisted doping, we dramatically increase/decrease the concentrations of acceptors/donors under positive external voltage and successfully reduce the Fermi level at room temperature to 2.10, 0.52, and 0.88 eV above VBM for the MBE, MOCVD, and HVPE without intentional dopants. By doping the β-$Ga_2O_3$ with Mg, the $E_{FR}$ can be further reduced.

## Declaration of competing interest

The authors declare that they have no known competing financial interests or personal relationships that could have appeared to influence the work reported in this paper.

## Data availability

The data are available in the Supplementary Information attached to this paper.

## Acknowledgements

This work was supported by the fund of Guangdong Provincial Key Laboratory of Computational Science

and Material Design (Grant No. 2019B030301001), the Shenzhen Science and Technology Innovation Commission (No. JCYJ20200109141412308), and the National Foundation of Natural Science, China (No. 52273226). All the calculations were carried out on the Taiyi cluster supported by the Center for Computational Science and Engineering of Southern University of Science and Technology and also on The Major Science and Technology Infrastructure Project of Material Genome Big-science Facilities Platform supported by Municipal Development and Reform Commission of Shenzhen.

## Graphical Abstract

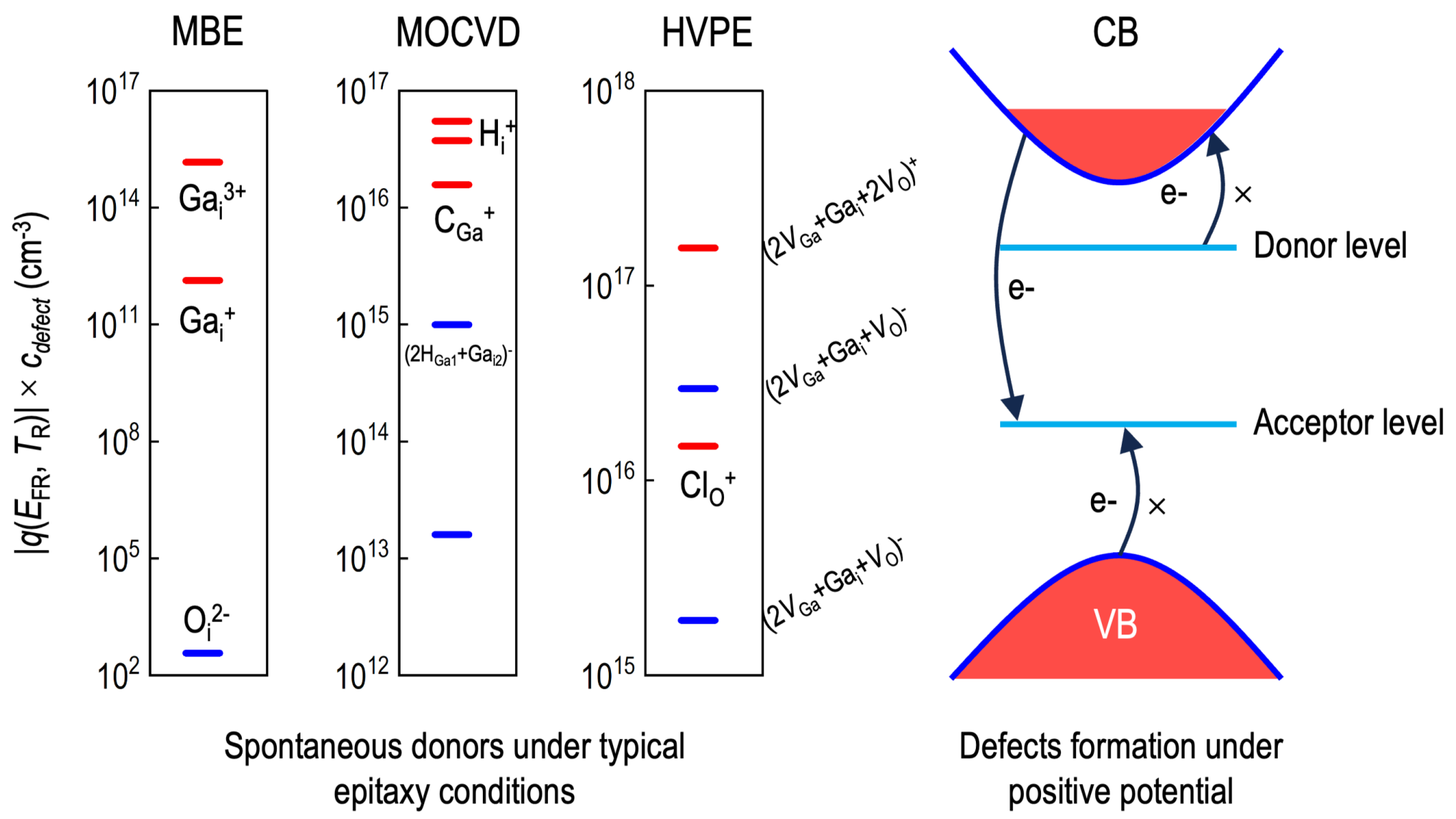

# Supplementary Information for "Spontaneous Donor Defects and Voltage–Assisted Hole Doping in Beta-Gallium Oxides under Multiple Epitaxy Conditions"

Chenxi Nie[1], Kai Liu[1], Chengxuan Ke[1], Xisong Jiang[2], Yifeng He[2], Yonghong Deng[1], Yanhua Yan[2,*], and Guangfu Luo[1,3,*]

[1]*Department of Materials Science and Engineering, Southern University of Science and Technology, Shenzhen 518055, China*

[2]*Shenzhen CAPCHEM Technology Co. Ltd., Shabo Tongfuyu Industry Zone, Pingshan District, Shenzhen 518118, China*

[3]*Guangdong Provincial Key Laboratory of Computational Science and Material Design, Southern University of Science and Technology, Shenzhen 518055, China*

**E-mail: yanyh@capchem.com, luogf@sustech.edu.cn*

## I. Tests of Computational Parameters for the HSE+U Method

To identify the optimal parameters for the HSE+U method, we examine the dependence of band gap and formation enthalpy of β-$Ga_2O_3$ as a function of $U_{eff}$ and the Hatree-Fock mixing parameter $\alpha$ in the HSE functional. As shown in Fig. S1, the parameters in the overlapping region between $E_g^{expt} \pm 0.1$ eV and $\Delta H^{expt} \pm 0.1$ eV exhibit the best overall performance. Given that the widely used HSE06 functional ($\alpha$ = 0.25) falls in this overlapping region, we choose $\alpha$ = 0.25 and $U_{eff}$ = 2.7 eV for our calculations.

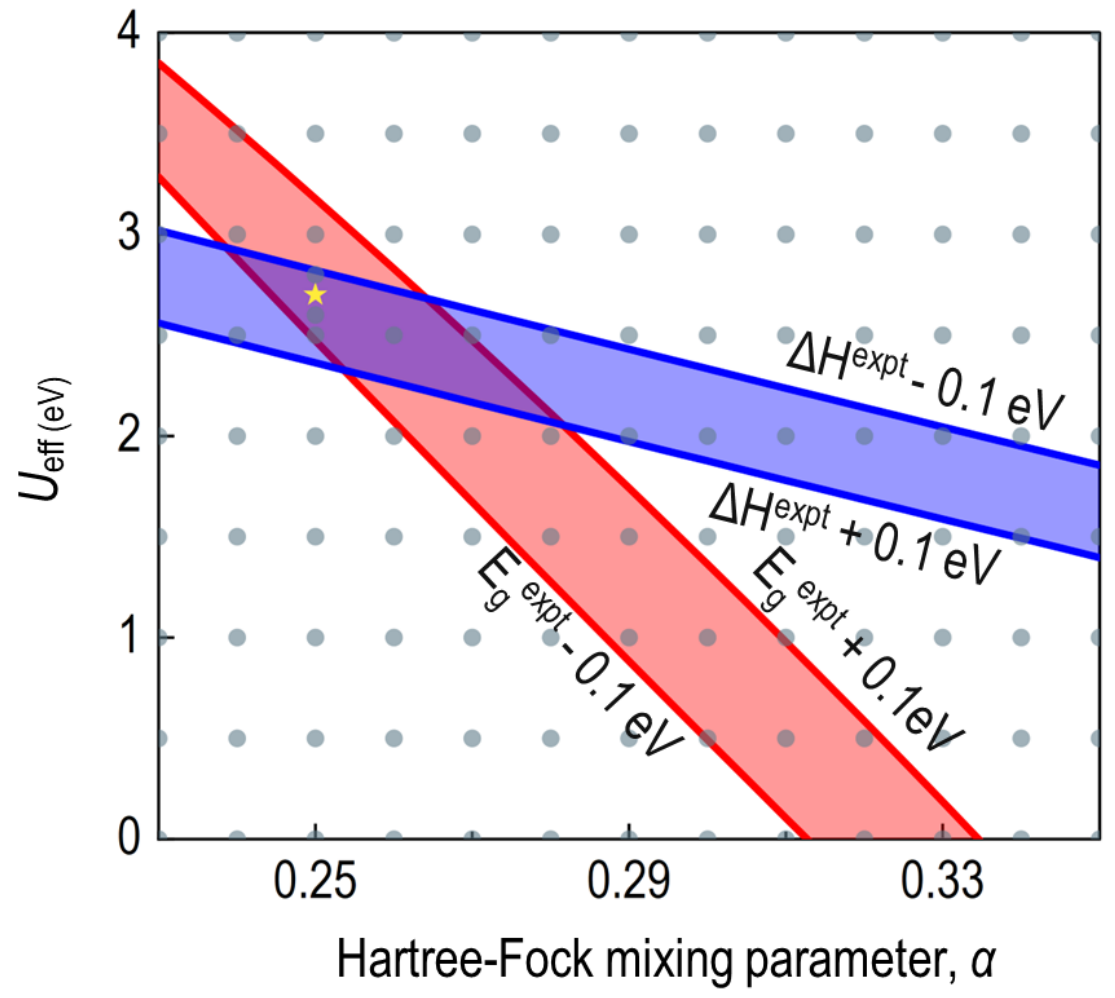

Figure S1. Predicted band gap and formation enthalpy of β-$Ga_2O_3$ as a function of $U_{eff}$ and the Hatree-Fock mixing parameter $\alpha$ in the HSE functional. Yellow star corresponds to HSE06 functional ($\alpha$ = 0.25) and $U_{eff}$ = 2.7 eV; gray dots indicate the examined parameters. The experimental band gap near 0 K is 4.86 eV[1, 2] and the experimental formation enthalpy is -11.29 eV obtained at 298.15 K and 1 bar[3, 4].

Table S1 indicates that HSE06+U ($U_{eff}$ = 2.7 eV) gives the best overall predictions on lattice constants, bandgap, and formation enthalpy of β-$Ga_2O_3$, with errors no greater than 1.8% of the experimental values. By contrast, the HSE method with $\alpha$ = 0.32, widely used in previous studies of β-$Ga_2O_3$[5-7], significantly underestimate the formation enthalpy by ~0.77 eV, despite its satisfactory prediction of the band gap.

Table S1. Comparison of lattice constant, band gap, and formation enthalpy of β-$Ga_2O_3$ among predictions by different computational methods and experimental values.

| | HSE06 | HSE ($\alpha$ = 0.32) | HSE06+U, $U_{eff}$ = 2.7 eV | Experiment |
|---|---|---|---|---|
| $a$ (Å) | 12.273 | 12.193 | 12.053 | 12.228† |
| $b$ (Å) | 3.043 | 3.034 | 2.985 | 3.038† |
| $c$ (Å) | 5.798 | 5.790 | 5.710 | 5.803† |
| $\beta$ (°) | 103.678 | 103.915 | 103.796 | 103.843† |
| $E_g$ (eV) | 4.19 | 4.86 | 4.81 | ~4.86§ |
| $\Delta H$ (eV) | -10.26 | -10.52 | -11.33 | -11.29‡ |

†The lattice constants are extrapolated to 0 K based on experimental results in the temperature range of 4–1200 K[8-10].

§The band gap is an average of the GW prediction of 4.80 eV at 0 K[2] and the experimental value of 4.92 eV at 20 K[1].

‡The experimental formation enthalpy was obtained under 298.15 K and 1 bar[3, 4].

## II. Temperature Dependence of Band Edges of β-$Ga_2O_3$

To obtain the temperature dependence of VBM energy of β-$Ga_2O_3$, we carry out molecular dynamics (MD) simulations at the PBE level under 400, 800, 1200, 1600, and 2000 K. Given the weak thermal expansion of β-$Ga_2O_3$[8-10], we adopt the NVT ensemble and a 12.05 × 8.95 × 11.42 Å$^3$ supercell predicted by HSE06+U geometry optimization. After reaching thermal equilibrium, six MD structures are sampled to obtain the average VBM and CBM energies at the HSE06+U level (Fig. S2a) through single-point energy calculations. These MD structures involve the dominant thermal effects, namely lattice vibrations[11], through a "frozen phonons" manner. The temperature dependence of band gap in Fig. S2b are in the general trend with the available experiment values[1, 12-15], which currently exhibit deviations at the level of ~0.3 eV.

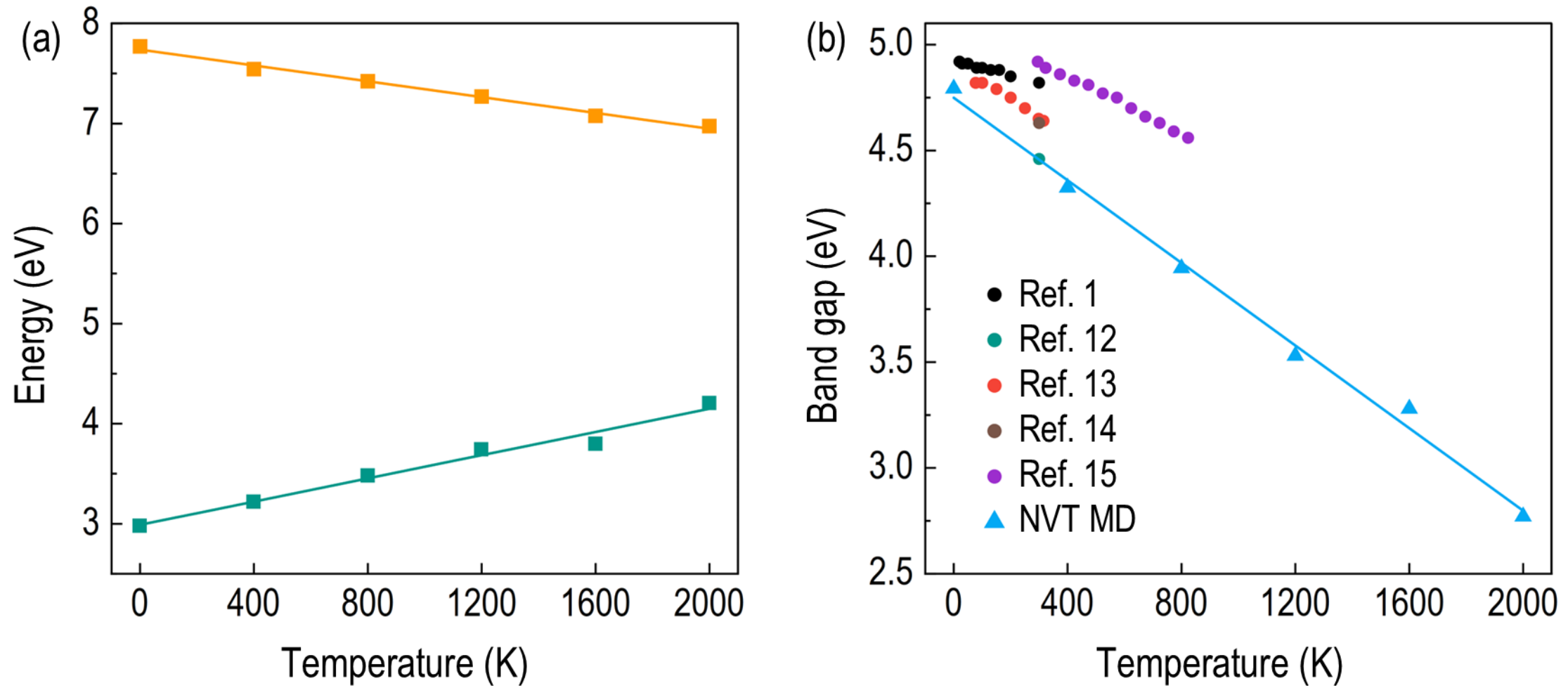

Figure S2. Temperature dependence of (a) band edges and (b) band gap for β-$Ga_2O_3$ from 0 to 2000 K. Available experimental band gaps[1, 12-15]are included for comparison.

## III. Assumption of Frozen Defect Concentrations during Cooling Process

In our model, we have used an assumption that defect concentrations do not significantly change during cooling process from growth temperature to room temperature, as can be justified with the following analyses. The mean squared displacement (MSD) of a defect is described by Eqn. (S1.1-S1.2)[16, 17], where $D$ is diffusion coefficient, $t$ is diffusion time, $\Gamma$ and $f$ are geometric and correlation factor, $a$ is the hopping distance, $v_0$ is the attempt frequency, $T$ is temperature, $E_f$ and $E_b$ are defect formation energy and diffusion barrier, respectively.

$$MSD = \sqrt{6Dt} \tag{S1.1}$$

$$D = \Gamma f a^2 v_0 e^{-\frac{E_f + E_b}{k_B T}} \tag{S1.2}$$

Typically, $\Gamma$*$f$ is at the level of ~10, $a$ is ~3.5 Å, attempt frequency is ~5 × $10^{12}$, $E_f$ is >2.3 eV in this study (Fig. 2 in the main text), $E_b$ is >0.28 eV (a lower limit based on the diffusion barrier of the very mobile defect $H_i$[18]). Assuming the sample is linearly cooled from growth temperature to 300 K within 5 minutes, the MSD during the cooling process is calculated to be less than 5.3 × $10^{-7}$, 2.9 × $10^{-8}$, and 8.2 × $10^{-10}$ m for the HVPE, MOCVD and MBE, respectively, according to Eqn. (S2.1-S2.3). These MSD values are small and noticeably less than the typical sample dimensions. Therefore, it is reasonable to assume that most defects would not migrate to the sample surface or annihilate with other defects as long as the cooling process is sufficiently fast.

$$MSD = \int_0^{300} \sqrt{\frac{3D}{2t}} dt \tag{S2.1}$$

$$D(t) = \Gamma f a^2 v_0 e^{-\frac{E_f}{k_B T_{growth}}} e^{-\frac{E_b}{k_B T(t)}} \tag{S2.2}$$

$$T(t) = \frac{300 - T_{growth}}{300} t + T_{growth} \tag{S2.3}$$

Here, $T_{growth}$ is growth temperature, which is 1273, 1023, and 823 K for HVPE, MOCVD and MBE, respectively.

Moreover, we have examined the concept of assuming all defect concentrations are in thermal equilibrium at 300 K. Taking MBE method as an example, adopting this approach yields results with the Fermi level positioned close to middle gap (~2.6 eV above valance band maximum) and an electron concentration around $10^{-7}$ $cm^{-3}$, which starkly deviates from the spontaneous n-type doping phenomenon observed in experiments. Such a low free carrier concentration originates from the low defect concentrations (< $10^{-6}$ $cm^{-3}$) at room temperature, as demonstrated by Eqn. S3. By contrast, the results derived from our current assumption give results aligning well with the experimental findings (Fig. 3 in the main text).

$$C_{defect} = C_{site} e^{\frac{-E_f}{k_B T}}, \tag{S3}$$

where $c_{site}$ is the maximum site concentration for a defect in crystal, $E_f$ is defect formation energy, and $T$ is temperature."

## IV. Selection of Defects to Investigate

For intrinsic point defects, we consider 10 vacancies and antisite defects, namely, $V_{O1}$, $V_{O2}$, $V_{O3}$, $V_{Ga1}$, $V_{Ga2}$, $Ga_{O1}$, $Ga_{O2}$, $Ga_{O3}$, $O_{Ga1}$, and $O_{Ga2}$. Also, we include 4 relatively stable Ga and O interstitials, namely, $Ga_{i1}$, $Ga_{i3}$, $O_{i1}$, and $O_{i3}$, based on the PBE results (Fig. S3a). For complex defects made of two-point defects, we exclude those related to the unstable $Ga_O$ and $O_{Ga}$ (Fig. S4), those made of point defects with the same charge, and those that transform automatically to single points. Eventually, we select 17 intrinsic defects for HSE06+U calculations: $V_{O1}$, $V_{O2}$, $V_{O3}$, $V_{Ga1}$, $V_{Ga2}$, $Ga_{O1}$, $Ga_{O2}$, $Ga_{O3}$, $O_{Ga1}$, $O_{Ga2}$, $Ga_{i1}$, $Ga_{i3}$, $O_{i1}$, $O_{i3}$, $Ga_{i3}$+$O_{i1}$, $V_{O2}$+$O_{i1}$, and $V_{O2}$+$V_{Ga2}$.

For impurity point defects, we examine the substitutional defects at both Ga and O sites and interstitial defects for C, H, and Cl impurities, based on previous experimental observations (Table 1). Given that $C_{Ga1}$ is more stable than $C_{Ga2}$[19], $H_{O1}$ is more stable than $H_{O2}$ and $H_{O3}$, and $Cl_{O1}$ is more stable than $Cl_{O2}$ and $Cl_{O3}$[20], we calculate 17 point impurity defects using PBE method: $C_{Ga1}$, $C_{O1}$, $C_{i1}$, $C_{i2}$, $C_{i3}$, $H_{Ga1}$, $H_{Ga2}$, $H_{O1}$, $H_{i1}$, $H_{i2}$, $H_{i3}$, $Cl_{Ga1}$, $Cl_{Ga2}$, $Cl_{O1}$, $Cl_{i1}$, $Cl_{i2}$, and $Cl_{i3}$. The PBE results in Fig. S3(b-d) reveal 6 impurity defects, namely, $C_{Ga1}$, $H_{Ga1}$, $H_{i1}$, $H_{i2}$, $Cl_{Ga1}$, and $Cl_{O1}$, with relatively low formation energies and thus are examined further with HSE06+U method.

For impurity-related complex defects, we apply the same principles as intrinsic complex defects and calculate 12 ones using PBE method: $C_{Ga1}$+$V_{Ga2}$, $C_{Ga1}$+$O_{i1}$, $H_{Ga1}$+$C_{Ga1}$, $H_{i2}$+$V_{Ga2}$, $H_{Ga1}$+$H_{i2}$, $H_{Ga1}$+$V_{O2}$, $H_{i2}$+$O_{i1}$, $H_{Ga1}$+$Ga_{i3}$, $Cl_{O1}$+$V_{Ga2}$, $Cl_{Ga1}$+$V_{O2}$, $Cl_{O1}$+$O_{i1}$, and $Cl_{Ga1}$+$Cl_{O1}$. Based on the PBE results in Fig. S3(e-g), we choose 4 impurity-related defects: $C_{Ga1}$+$V_{Ga2}$, $H_{Ga1}$+$C_{Ga1}$, $H_{Ga1}$+$H_{i2}$, and $Cl_{O1}$+$V_{Ga2}$ for HSE06+U calculations.

Besides the above 27 ones, we also examine 7 additional defects proposed in literature, including $V_{Ga1}$+$V_{Ga2}$+$Ga_{i1}$[21, 22], 2$V_{Ga1}$+$Ga_{i2}$[21, 22], 2$V_{Ga1}$+$Ga_{i3}$[21, 22], 2$V_{Ga1}$+$Ga_{i2}$+$V_{O1}$[5, 23], 2$V_{Ga1}$+$Ga_{i2}$+2$V_{O1}$[5], 2$V_{Ga1}$+$Ga_{i3}$+$V_{O3}$[5], and 2$H_{Ga1}$+$Ga_{i2}$[24]. The first three defects consist of two gallium vacancies and one interstitial gallium atom, which were first predicted through first-principles calculations[21] and later confirmed with high resolution scanning transmission electron microscopy[22]. The fourth to sixth defects involve an additional oxygen vacancy relative to the first three and were predicted through first-principles calculations[5]. The seventh defect, 2$H_{Ga1}$+$Ga_{i2}$, was identified through a combination of experimental and computational efforts[24]. Note that this defect was named differently in reference 24. In total, we calculate 34 defects using the HSE06+U method.

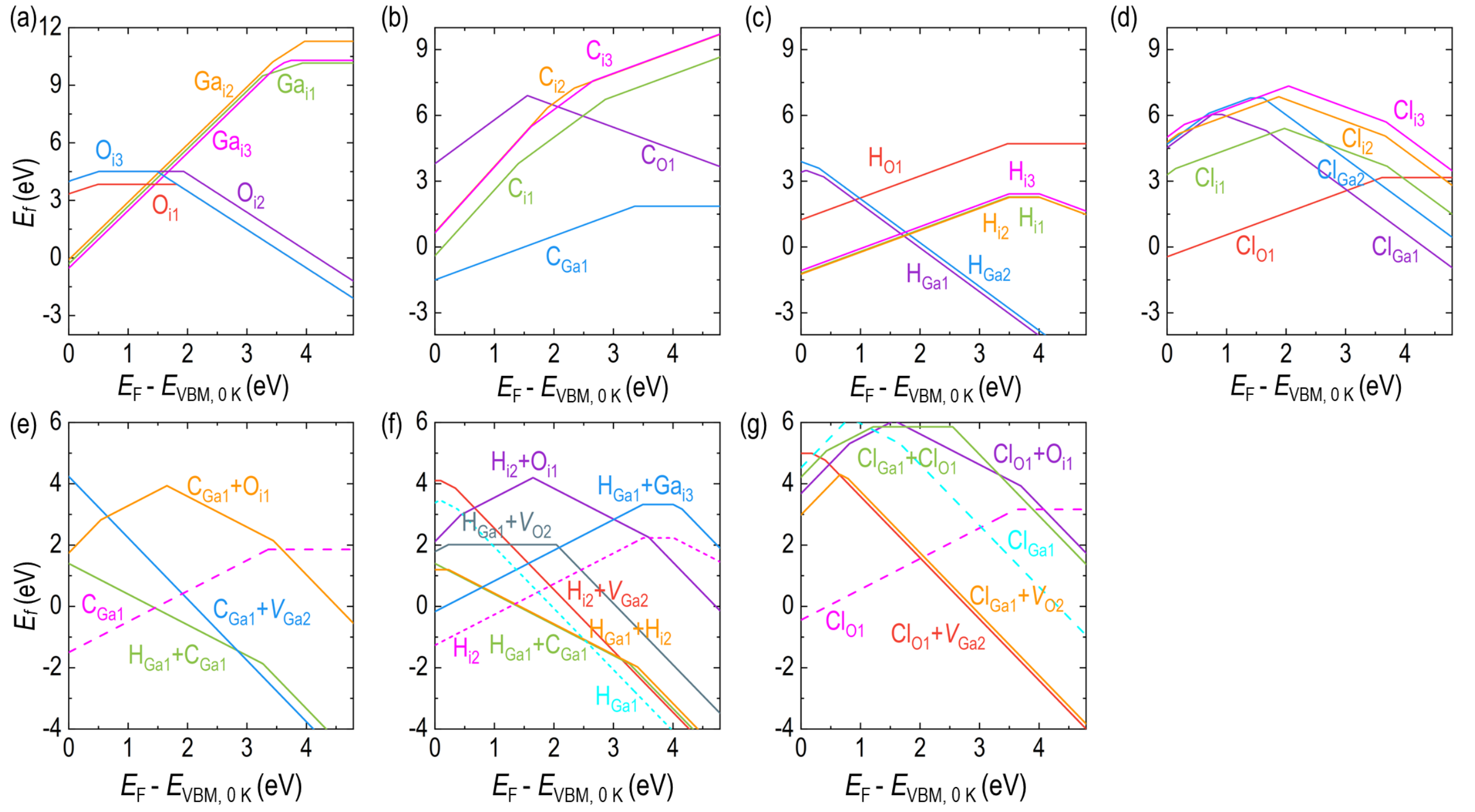

Figure S3. Defect formation energies of (a) Ga and O interstitials, and point defects related to impurities (b) C, (c) H, (d) Cl at the PBE level; defect formation energies for the (e) C-related, (f) H-related, and (g) Cl-related complex defects at the PBE level. The chemical potentials μ of O and Ga are based on the growth conditions of MOCVD for (a-c, e, f) and HVPE for (d, g); $\mu_C$ is the total energy of graphite per atom, $\mu_{Cl}$ is the total energy of $Cl_2$ gas per atom, $\mu_H = 0.5*(\mu_{H2O} - \mu_O)$.

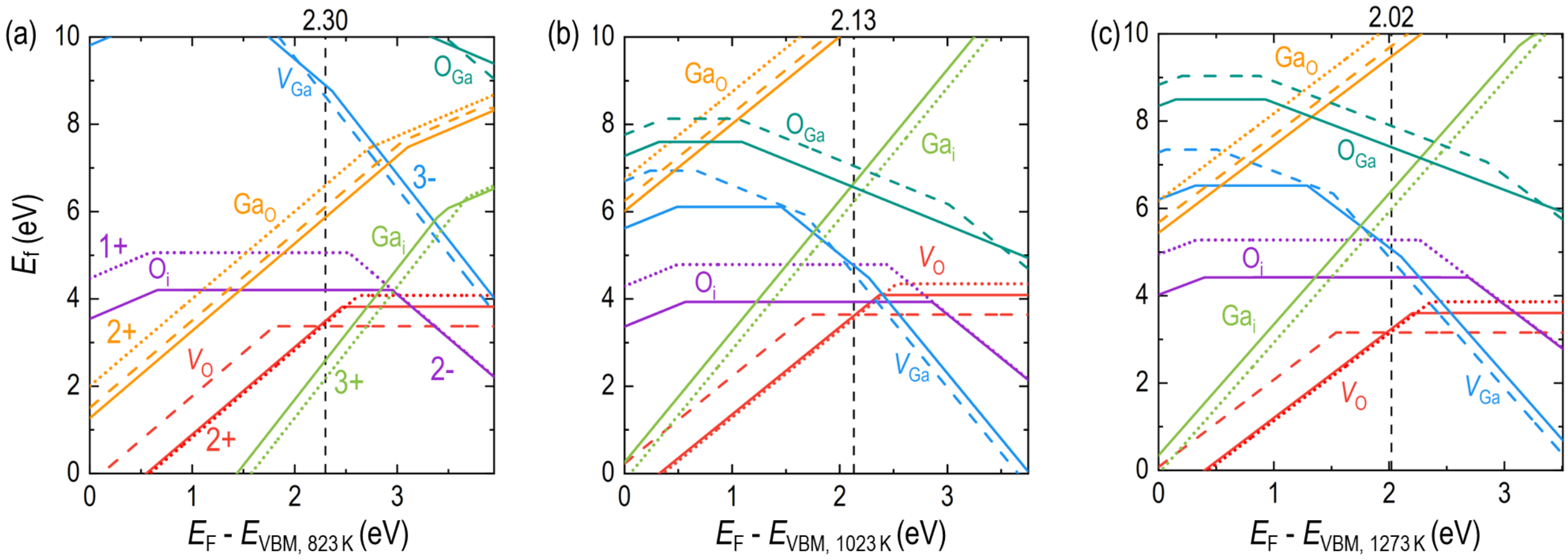

Figure S4. Formation energy of intrinsic point defects for (a)MBE, (b)HVPE, and (c)HVPE. The data here is same as that in Fig. 2(a, c, f) but shown with a wider energy range.

## V. MBE Method with Different Chemical Potentials

In MBE method, oxygen plasma or $O_3$ is sometimes utilized in the growth of metal oxides. Therefore, in addition to the pure $O_2$ source discussed in the main text, here we include two more conditions involving the use of $O_3$: one with pure $O_3$ and the other with a mixture of 50% $O_2$ and 50% $O_3$. As expected, the chemical potentials of oxygen in the two additional conditions are greater than that of pure $O_2$ (Table S2), owing to the elevated reactivity of $O_3$. Consequently, all defects involving oxygen are affected, while others remain unchanged.

The dominant defects, $Gi_3^{3+}$ and $Gi_1^{+}$, observed in the growth using pure $O_2$, persist as the dominant ones under the two new conditions (Fig. S5c, f, i). Although the concentrations of other major defects, such as $O_{i1}^{2-}$, $(Ga_{i3} + O_{i1})^{+}$, and $O_{i3}^{2-}$, are notably increased, they remain at very low levels of $< 10^9$ $cm^{-3}$. Therefore, from the viewpoint of defect thermodynamics, the influence of $O_3$ in determining the defects in MBE is minor.

Table S2. Three oxygen sources utilized for simulations of MBE growth of β-$Ga_2O_3$ at 823 K.

| Oxygen source | $O_2$ | 50%$O_2$ + 50%$O_3$ | $O_3$ |
|---|---|---|---|
| Total pressure (Pa) | ~$1.33 \times 10^{-3}$ | ~$1.33 \times 10^{-3}$ | ~$1.33 \times 10^{-3}$ |
| $\mu_O$ (eV) | -8.37 | -7.89 | -7.37 |

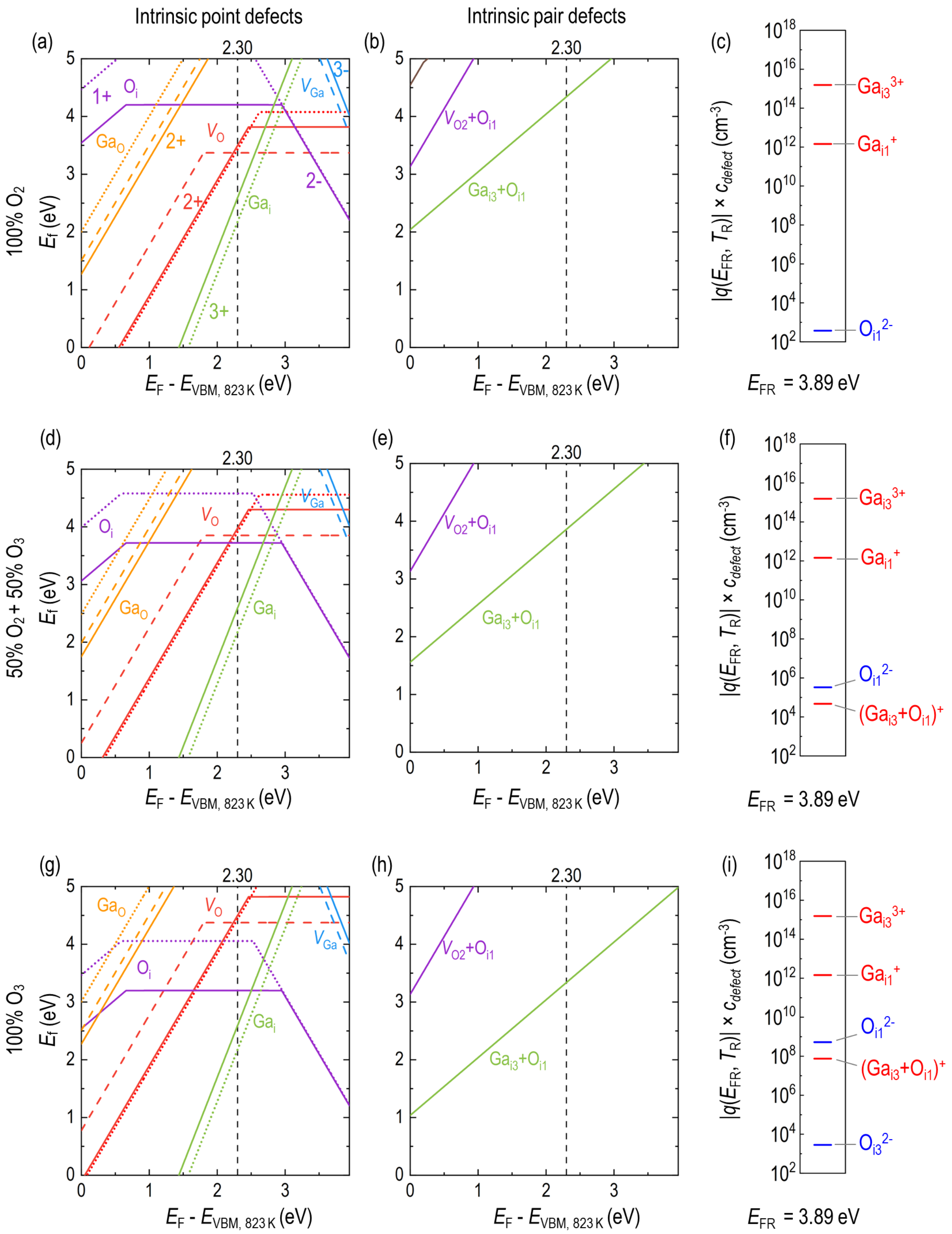

Figure S5. Defects formation energy for MBE-grown β-$Ga_2O_3$ using (a-b) pure $O_2$, (d-e) a mixture of 50% $O_2$ and 50% $O_3$, and (g-h) pure $O_3$ as oxygen sources. The other conditions remain the same as listed in Table 1 of the main text. Vertical dash line indicates the self-consistent Fermi level $E_{FG}$. Defect charge states are labeled besides the slopes in panel (a). The solid, dashed, and short curves represent defects occupying the first, second, and third inequivalent positions in β-$Ga_2O_3$ (see Fig. 2i in the main text), respectively. (c, f, i) Charge concentrations of defects and self-consistent Fermi level ($E_{FR}$) at room temperature for the corresponding condition.

We also examine the intrinsic defects under the Ga-rich and O-rich conditions at the growth temperature of MBE (823 K) for comparison. Our findings indicate that the Ga-rich and O-rich conditions exhibit dominant defects different from the MBE methods and show a weak n- and p-type doping, respectively (Fig. S6).

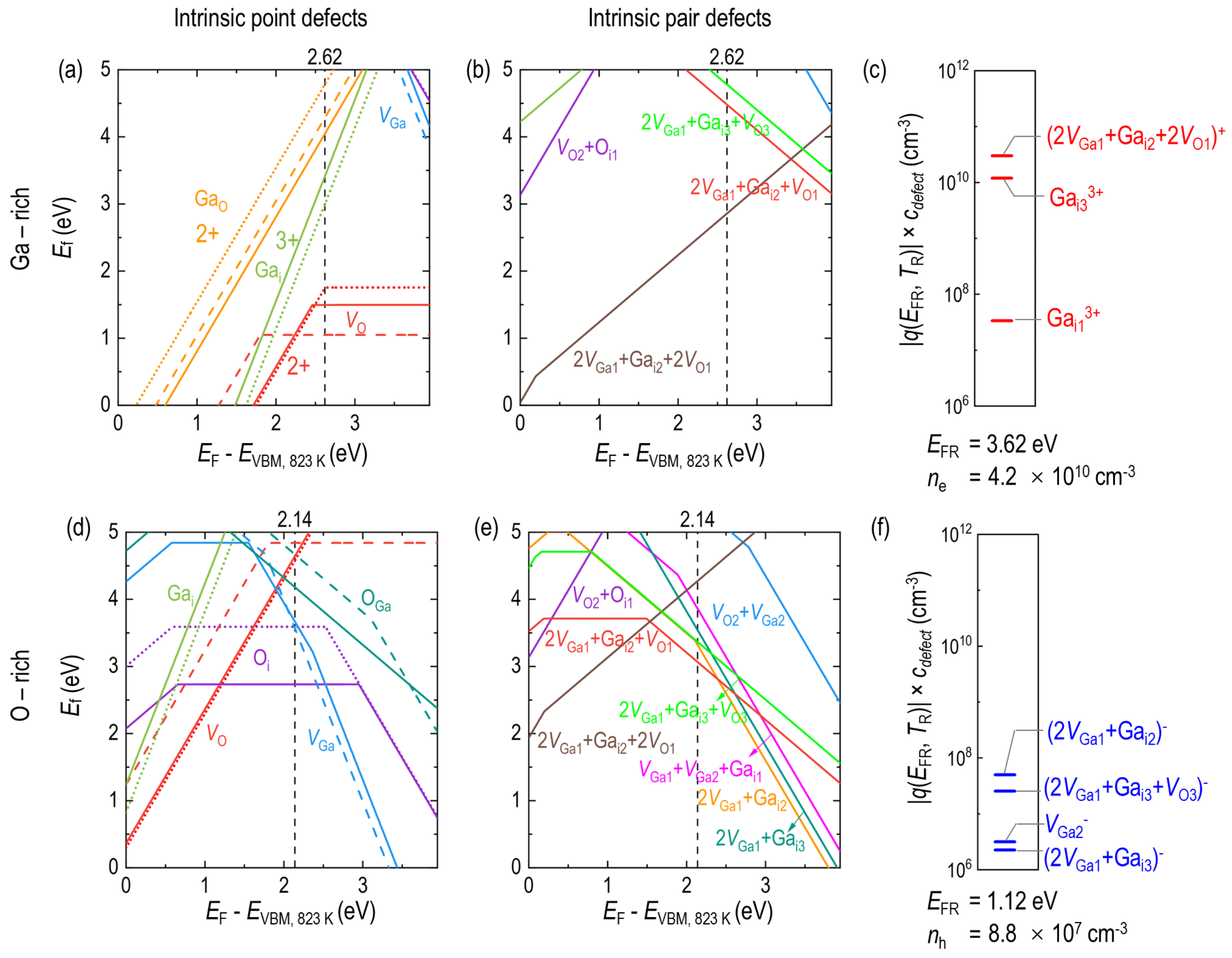

Figure S6. Defects formation energy for intrinsic defects under the (a-b) Ga-rich and (d-e) O-rich conditions at 823 K. Vertical dash line indicates the self-consistent Fermi level $E_{FG}$. Defect charge states are labeled besides the slopes in panel (a). The solid, dashed, and short curves represent defects occupying the first, second, and third inequivalent positions in β-$Ga_2O_3$, respectively. (c, f) Charge concentrations of defects, self-consistent Fermi level ($E_{FR}$) and free electron concentration ($n_e$) or free hole concentration ($n_h$) at room temperature.

## VI. Comparison of Predicted Charge Transition Levels with Experiments

To establish connections between our predictions and available experimental data, we compare the measured defect energy levels in β-$Ga_2O_3$ samples grown via MBE, MOCVD, and HVPE[25-31] with those of the dominant defects predicted in this study (Fig. 3 in the main text). As depicted in Fig. S7, the experimental defect level at 0.56 eV[26] in an unintentionally doped MBE-grown sample aligns well with the energy level of the dominant defect $Ga_{i3}^{3/1}$ at 0.62 eV of this work. In the case of MOCVD, the experimental energy levels around 4.4 eV[27, 28] could be associated with the dominant defect $2H_{Ga1}+Ga_{i2}$. Concerning HVPE, the experimental energy level at 4.2 eV[31] correspond to those of three dominant defects: $2V_{Ga1}+Ga_{i2}+V_{O1}$, $2V_{Ga1}+Ga_{i3}+V_{O3}$, and $2V_{Ga1}+Ga_{i2}+2V_{O1}$.

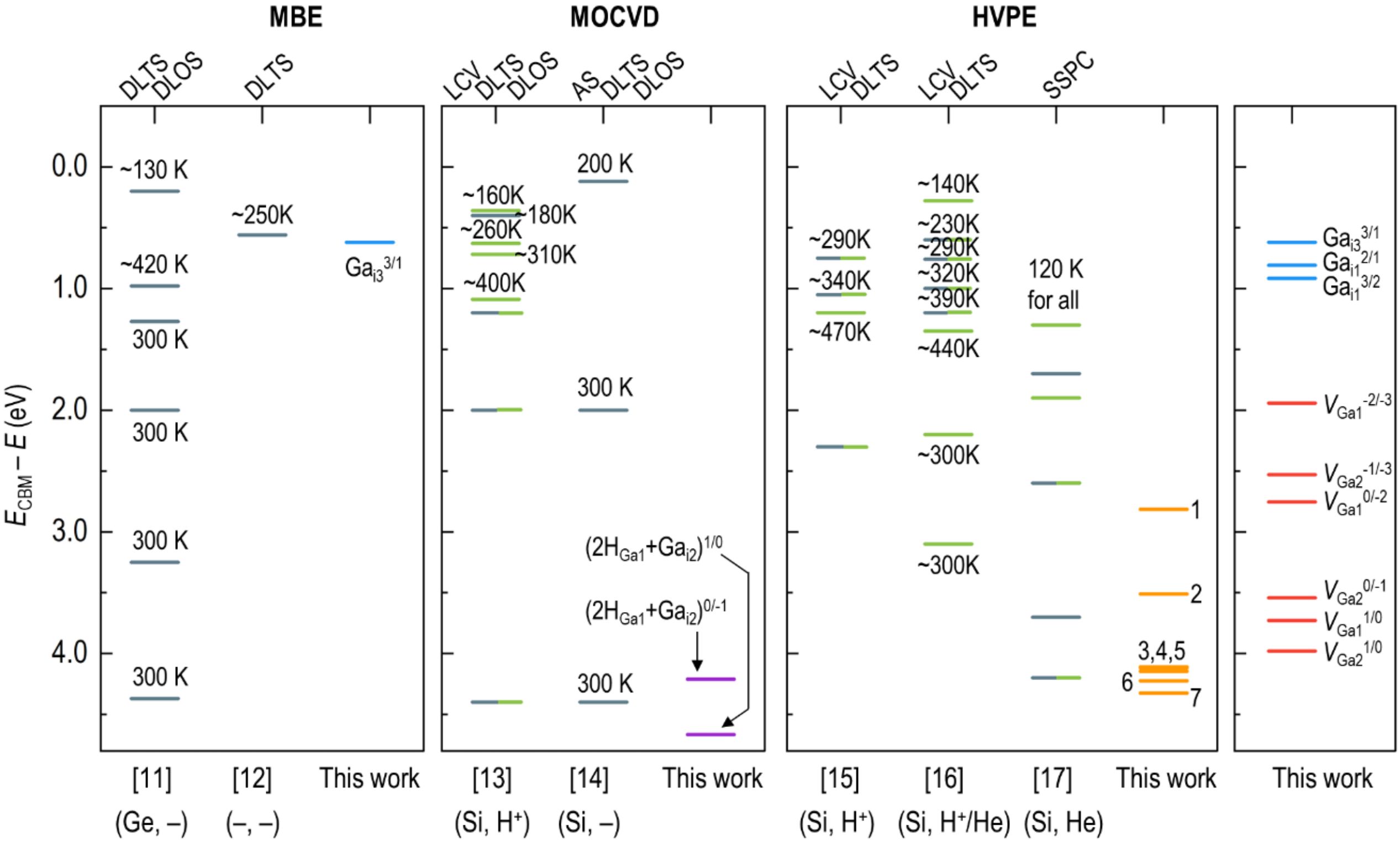

Figure S7. Comparison between experimental defect energy levels and the predicted defect energy levels of the dominant defects for the MBE, MOCVD, and HVPE experiments. Experimental defect energy levels with estimated concentrations less than $10^{14}$ $cm^{-3}$ were excluded for clarity. The temperature of measuring a defect energy level is labeled whenever possible, and our simulated results are referenced to CBM at 0 K. Labels above the top axis indicate the methods employed to measure the defect energy levels: DLTS for deep-level transient spectroscopy, DLOS for deep level optical spectroscopy, LCV for lighted capacitance–voltage measurement, AS for admittance spectroscopy, and SSPC for steady-state photocapacitance. The first and second values in each parenthesis below the bottom axis, respectively, indicate the intentional dopant and the species used to irradiate β-$Ga_2O_3$. Experimental defect energy levels shown as complete green bars indicate ones appearing after irradiation, while those with half green bars represent ones with enhanced intensity after irradiation. The numbers besides the predicted energy levels in HVPE have the following meaning: 1, 4, and 7 for the 0/-1, 1/0, and 2/1 energy level of $2V_{Ga1}+Ga_{i2}+V_{O1}$; 2, 5, and 6 for the 0/-1, 1/0, and 2/1 energy level of $2V_{Ga1}+Ga_{i3}+V_{O3}$; 3 for 2/1 energy level for $2V_{Ga1}+Ga_{i2}+2V_{O1}$.

Furthermore, certain experiments intentionally irradiated the samples with $H^+$ or He, which are expected to induce defects, such as $Ga_i$ and $V_{Ga}$, owing to the larger radius of gallium than oxygen. Our predicted energy levels of $Ga_i$ fall within the range of 0.62 – 0.92 eV, overlapping with the enhanced defect signals in the range of 0.36 – 1.1 eV. The predicted energy levels of $V_{Ga}$ spread in the range of 1.94 – 3.98 eV, the correlation of which with experimental defect levels is not straightforward.

In summary, the defect energy levels of the dominant defects predicted in this study can be correlated with the experimental data. However, discrepancies among experimental data underscore the complexity, which could arise from the existence of impurities and variation in measurement temperature, the latter of which is known to impact the positions of defect levels [Phys. Rev. B 105, 115201 (2022)].

## VII. Impacts of Growth Temperature on Defects Formation

In addition to the growth temperature of MBE at 823K and MOCVD at 1023K, as discussed in the main text, here we include more growth temperature for the two methods. Since the widely used growth temperature (1273 K) for the HVPE method yields high growth rate, mirrorlike surface, and high structural quality[32-35], further exploration on other growth temperature seems unnecessary. Table S3 and S4 display the growth conditions for the MBE and MOCVD, respectively. The results in Fig. S8 and S9 indicate that the dominant defect types remain the same under different temperatures, while the defect and carrier concentrations slightly increase or decrease depending on the growth techniques.

Table S3. Growth conditions and chemical potentials at two temperatures for our MBE simulation.

| $T_G$ (K) | 823 | 923 |
|---|---|---|
| $P_{O_2}$ (Pa) | ~$1.33 \times 10^{-3}$ | |
| Ga source, $P_{Ga}$ (Pa) | Ga vapor, ~$2.0 \times 10^{-4}$ | |
| $\mu_O$ (eV)† | -8.37 | -8.57 |
| $\mu_{Ga}$ (eV)‡ | -3.41 | -3.78 |

†The translational, vibrational, and rotational entropies are included under experimental conditions.

‡ The chemical potential of $\mu_{Ga}$ is calculated directly from the pressure and temperature of gallium vapor.

Table S4. Growth conditions and chemical potentials at two growth temperatures for our MOCVD simulation. Impurity concentrations are extracted from experiments[36].

| $T_G$ (K) | 1023 | 1153 |
|---|---|---|
| $P_{O_2}$ (Pa) | ~$6.08 \times 10^{3}$ | |
| Ga source, $P_{Ga}$ (Pa) | TEGa gas, 2.7-10 | |
| $\mu_O$ (eV)† | -8.10 | -8.28 |
| $\mu_{Ga}$ (eV)‡ | -7.69 | -7.58 |
| $C_{impurity}$ ($cm^{-3}$) | C: ~ $2.0 \times 10^{16}$<br>H: ~$1.2 \times 10^{17}$ | C: ~$1.5 \times 10^{16}$<br>H: ~$3.5 \times 10^{16}$ |
| §$\mu_{impurity}$ (eV) | $\mu_C$: -12.76<br>$\mu_H$: -6.22 | $\mu_C$: -12.98<br>$\mu_H$: -6.64 |

†The translational, vibrational, and rotational entropies are included under experimental conditions.

‡The chemical potential of $\mu_{Ga}$ is obtained using the equilibrium condition of $2\mu_{Ga} + 3\mu_{O} = \mu_{Ga2O3}$ under experimental conditions.

§The chemical potentials of impurities are determined in a way that the impurity concentrations are consistent with the experimental values.

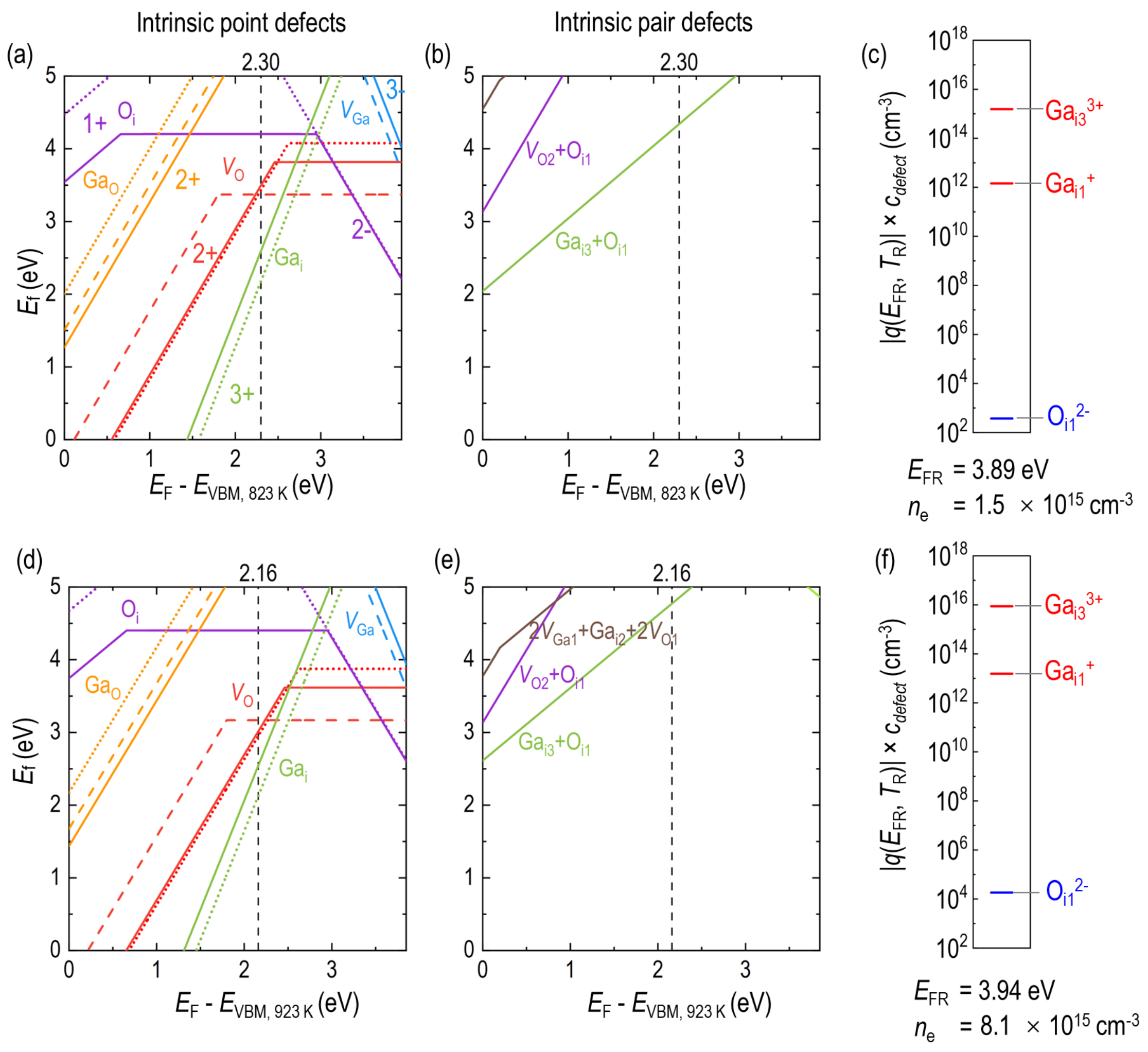

Figure S8. Defects formation energy for MBE at temperatures of (a-b) 823 and (d-e) 923 K. Vertical dash line indicates the self-consistent Fermi level $E_{FG}$. Defect charge states are labeled besides the slopes in panel (a). The solid, dashed, and short curves represent defects occupying the first, second, and third inequivalent positions in β-$Ga_2O_3$, respectively. (c, f) Charge concentrations of defects, self-consistent Fermi level ($E_{FR}$) and free electron concentration ($n_e$) at room temperature.

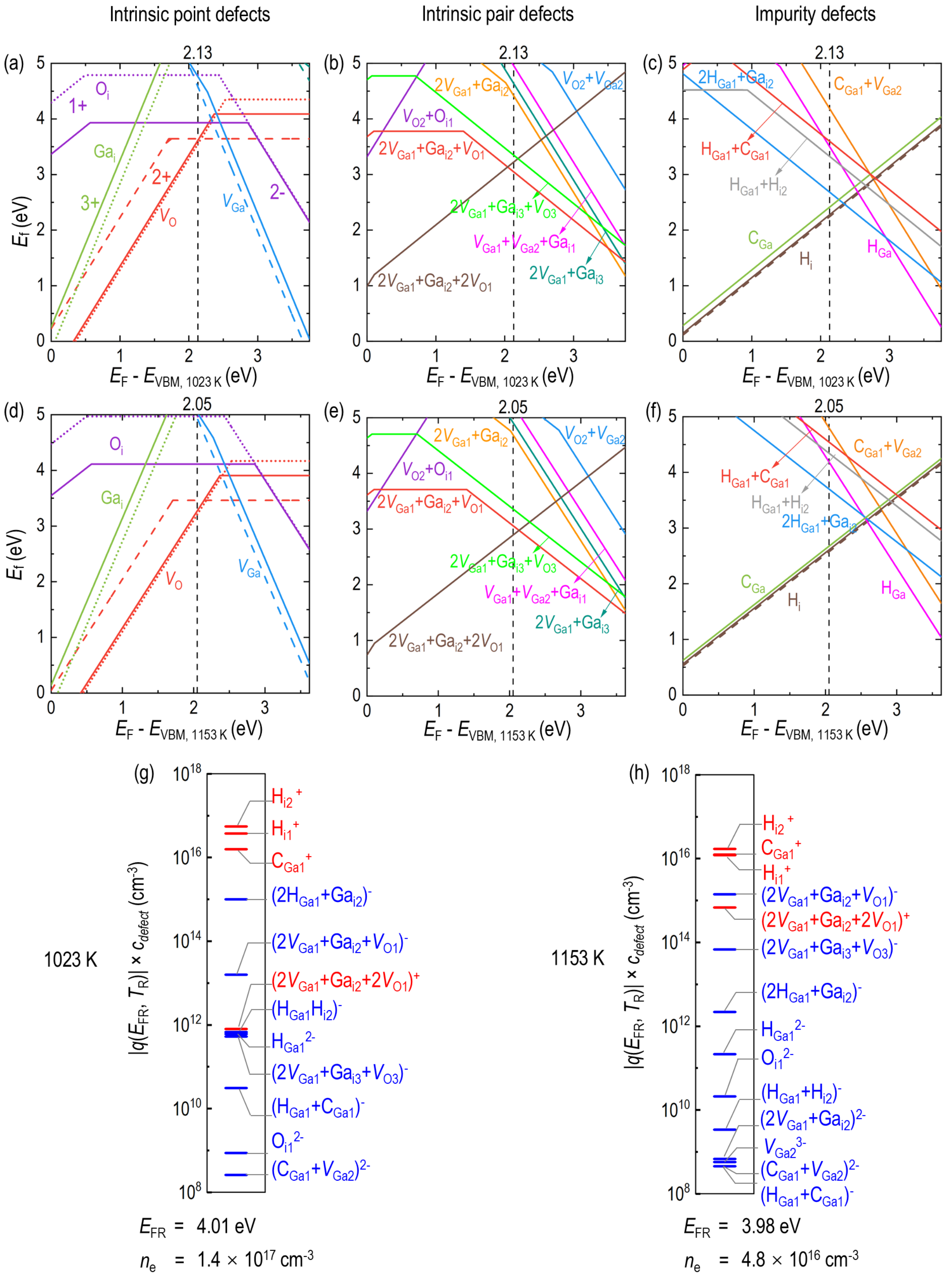

Figure S9. Defects formation energy for MOCVD at temperatures of (a-c) 1023 and (d-f) 1153 K. Vertical dash line indicates the self-consistent Fermi level $E_{FG}$. Defect charge states are labeled besides the slopes in panel (a). The solid, dashed, and short curves represent defects occupying the first, second, and third inequivalent positions in β-$Ga_2O_3$, respectively. (c, f) Charge concentrations of defects, self-consistent Fermi level ($E_{FR}$) and free electron concentration ($n_e$) at room temperature.

## VIII. Comparison of Ionization Ratio for Acceptors in β-$Ga_2O_3$ and Si

The ionization of an acceptor is both inversely proportional to the ionization energy level and proportional to the effective hole mass and band gap, as demonstrated in Eqn. S4[37]. Although the ionization energy level of 0.39 eV for $Mg_{Ga}^{0/-}$ at 300 K is notably higher than that of typical dopants in traditional semiconductors, such as ~0.05 eV for $B^{0/-}$[38] in Si, β-$Ga_2O_3$ has relatively flatten valance bands with a heavy effective hole mass of ~7.6 $m_e$ and a wide band gap of ~4.5 eV at 300 K. As depicted in the Fig. S10, β-$Ga_2O_3$ can support noticeably higher ionization energy level than Si.

$$2\frac{(2\pi m_h^* k_B T)^{3/2}}{h^3} e^{\frac{-E_F}{k_B T}} - 2\frac{(2\pi m_e^* k_B T)^{3/2}}{h^3} e^{\frac{E_F - E_g}{k_B T}} - \frac{N_a}{1+g_a e^{\frac{E_a - E_F}{k_B T}}} = 0 \quad \text{(S4)}$$

Here, $m_h^*$ is effective hole mass, $m_e^*$ is effective electron mass, $T$ is temperature, $E_F$ is Fermi level relative to the valance band maximum, $E_g$ is band gap, $N_a$ is acceptor concentration, $E_a$ is acceptor ionization energy level, and $g_a$ is the acceptor degeneracy factor, which is 2 or 4 for most semiconductors.

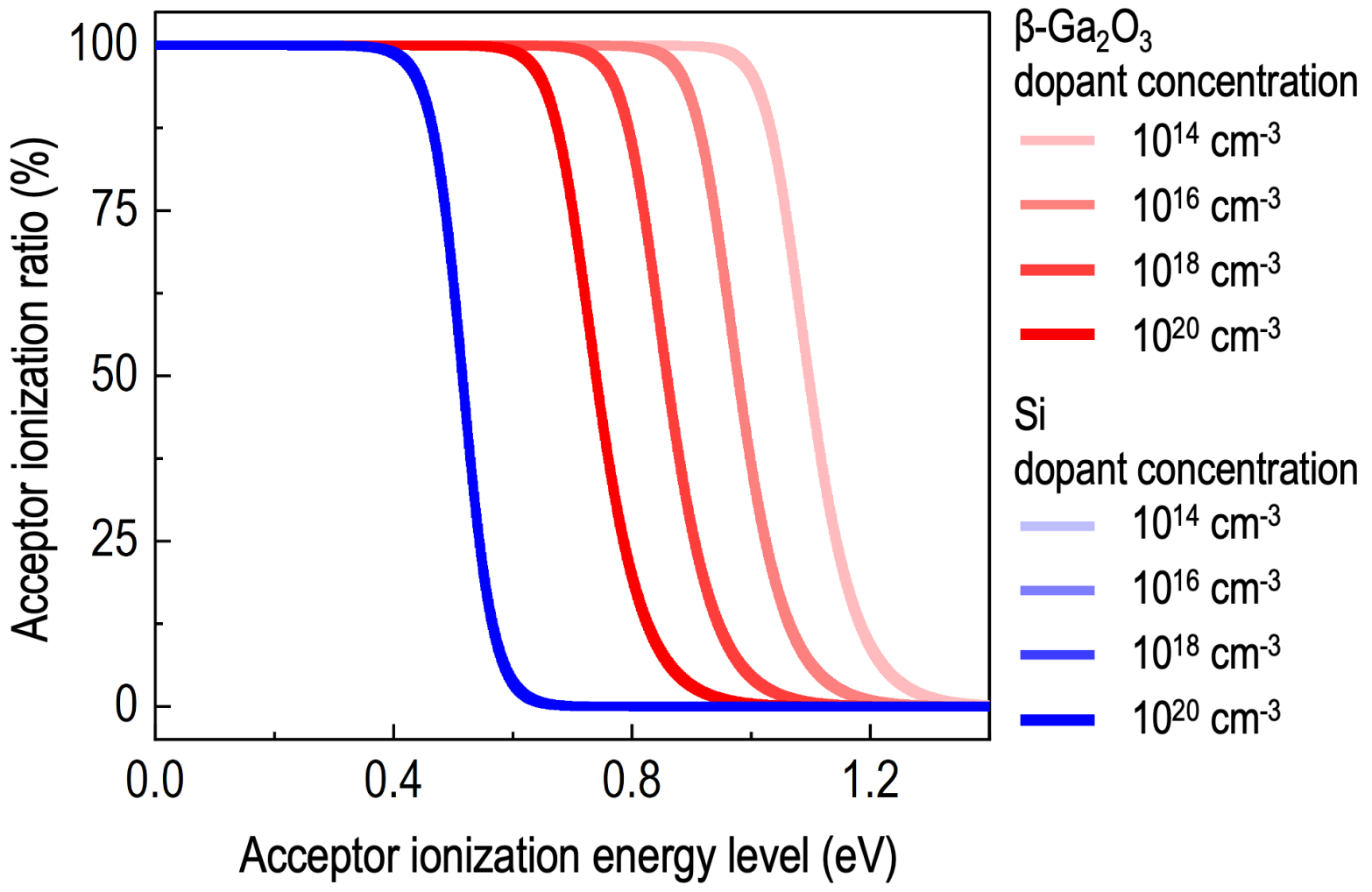

Figure S10. Ionization ratio for acceptors in β-$Ga_2O_3$ and Si at 300 K with a dependence of acceptor ionization energy level and acceptor concentration. The respective effective electron and hole masses used here are 0.28 and 7.6 $m_e$, values fitted from density of states, for β-$Ga_2O_3$ or 1.09 and 1.15 $m_e$ for Si[39]; the acceptor degeneracy factor is set to 4.

## IX. Charge Transition Level of $Mg_{Ga2}^{0/-}$

Table S5 compares the charge transition levels of $Mg_{Ga2}^{0/-}$ among experimental reports and theoretical predictions. The defect formation energies of $Mg_{Ga2}$ based on HSE06+U ($U_{eff}$ = 2.7 eV) and HSE ($\alpha$ = 0.32) are also provided in Fig. S11. It is evident that the predicted transition level of $Mg_{Ga2}^{0/-}$ is strongly influenced by the theoretical method. Given our analyses in Fig. S1 and the past superior performance of the HSE06+U method in II-VI semiconductors [40], we are inclined to believe that the prediction by HSE06+U ($U_{eff}$ = 2.7 eV) may be more accurate. The discrepancies among experimental results underscore the complexities in experiments, which often necessitate theoretical predictions for comparison in order to assign the defect origins.

Table S5 Summary of charge transition level of $Mg_{Ga2}^{0/-}$

| Method | $E(0/-)$ |
|---|---|
| HSE ($\alpha$ = 0.32) | $E_{VBM(0\ K)}$ + 1.06 eV[6]<br>$E_{VBM(0\ K)}$ + 1.05 eV[7]<br>$E_{VBM(0\ K)}$ + 1.03 eV [this work]<br>$E_{VBM(300\ K)}$ + 0.84 eV [this work] |
| HSE06+U ($U_{eff}$ = 2.7 eV) | $E_{VBM(0\ K)}$ + 0.58 eV [this work]<br>$E_{VBM(300\ K)}$ + 0.39 eV [this work] |
| Photoinduced current transient spectroscopy | $E_{VBM(370\text{-}410\ K)}$ + 1.05 eV[41] |
| Photoinduced electron paramagnetic resonance | $E_{VBM(130\ K)}$ + 1.2 eV[42] |
| Electron paramagnetic resonance | $E_{VBM(240\text{-}260\ K)}$ + 0.65 eV[43]<br>< $E_{VBM(255\ K)}$ + 0.7 eV[44] |

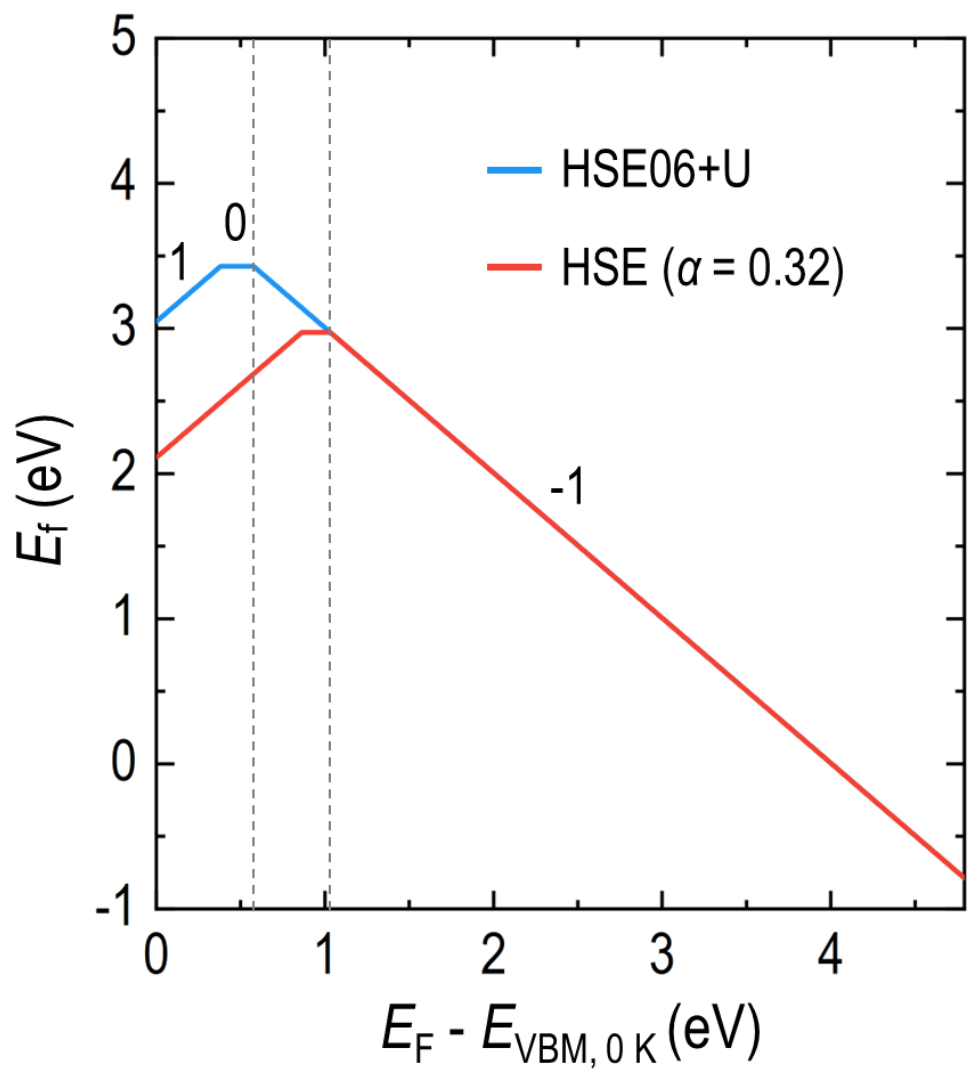

Figure S11. Formation energy of $Mg_{Ga2}$ based on HSE06+U ($U_{eff}$ = 2.7 eV) and HSE ($\alpha$ = 0.32). VBM energy at 0 K is used; $\mu_{Mg}$ and $\mu_{Ga}$ are set to zero for convenience, as does not impact the charge transition levels.